\documentclass[journal=jacsat,manuscript=article]{achemso}
\usepackage[version=3]{mhchem} 

\usepackage{siunitx}[=v2]
\usepackage{physics} 
\usepackage{chemformula}
\usepackage{textgreek}
\usepackage{float}

\usepackage{verbatim}
\usepackage{xcolor}

\SectionNumbersOn
\DeclareSIUnit\molar{M}
\DeclareSIUnit\photon{photon}
\DeclareSIUnit\OD{OD}

\usepackage{booktabs}
\usepackage{tabularx}
\usepackage{makecell}

\usepackage{xcolor}

\usepackage{soul}

\usepackage{textgreek}

\usepackage{graphicx}			
\DeclareGraphicsExtensions{.pdf,.png,.jpg} 

\usepackage{siunitx}		
\DeclareSIUnit\photon{photon}
\DeclareSIUnit\molar{M}
\DeclareSIUnit\Molar{M}
\DeclareSIUnit\rev{rev} 
\DeclareSIUnit\OD{OD}
\DeclareSIUnit\cal{cal} 
\DeclareSIUnit\calorie{cal} 
\DeclareSIUnit\lines{lines}

\usepackage{bm}

\DeclareUnicodeCharacter{2032}{'} 
\DeclareUnicodeCharacter{2212}{--} 

\author{Robert H. Gordon}
\affiliation{School of Mathematical and Physical Sciences, University of Sheffield, Sheffield, UK}

\author{James P. Pidgeon}
\affiliation{School of Mathematical and Physical Sciences, University of Sheffield, Sheffield, UK}

\author{Rachel C. Kilbride}
\affiliation{Department of Physics, University of Warwick, Coventry, UK}
\alsoaffiliation{XMaS, The UK X-ray Materials Science Facility, European Synchrotron Radiation Facility, Grenoble, France}

\author{Tingxiang Yang}
\affiliation{School of Biosciences, University of Sheffield, Sheffield, UK}

\author{Benjamin Stewart}
\affiliation{Department of Chemistry, University of Oxford, Oxford, UK}

\author{Yan Chen}
\affiliation{School of Mathematical and Physical Sciences, University of Sheffield, Sheffield, UK}

\author{Timothy Thornber}
\affiliation{School of Mathematical and Physical Sciences, University of Sheffield, Sheffield, UK}

\author{Peter Claronino}
\affiliation{School of Mathematical and Physical Sciences, University of Sheffield, Sheffield, UK}

\author{Graham J. Leggett}
\affiliation{School of Mathematical and Physical Sciences, University of Sheffield, Sheffield, UK}

\author{Claudia E. Tait}
\affiliation{Department of Chemistry, University of Oxford, Oxford, UK}

\author{Marie Rider}
\affiliation{Department of Physics, University of Bath, Bath, UK}

\author{Jenny Clark}
\email{jenny.clark@sheffield.ac.uk}
\phone{+44 (0)114 222 3526}
\affiliation{School of Mathematical and Physical Sciences, University of Sheffield, Sheffield, UK}

\title[An \textsf{achemso} demo]
  {NIR-probe transient absorption on rubrene microcavities demonstrates no changes to singlet fission by strong-coupling}

\abbreviations{TA, TMM, SF, TTA, NIR, DBR, ESA}

\begin{document}

\begin{tocentry}





\includegraphics[scale=1.0]{"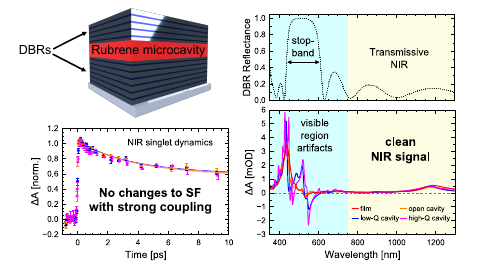"}

\end{tocentry}




\begin{abstract}\label{abstract}
Strong light-matter coupling has been proposed to control singlet-exciton fission, yet reported effects remain controversial due to the prevalence of cavity-induced optical artifacts in transient absorption measurements. Here, we investigate singlet fission in dielectric microcavities containing rubrene by probing in the near-infrared region, outside the cavity stopband, to eliminate mode-filtering artifacts. Across contrasting cavity quality factors ($Q = 90$ and $1150$) and a range of exciton-photon detunings ($+20$ to $-110$ meV), we find that the underlying kinetics are virtually identical to uncoupled films. Although minor spectral changes occur, transfer matrix simulations reveal these stem purely from mirror interference rather than modified photophysics. Transient electron paramagnetic resonance further confirms that high-spin triplet dynamics remain unperturbed. Together, these findings establish a widely applicable experimental framework for evaluating molecular photophysics in microcavities, revealing that singlet fission remains unaltered in strongly coupled rubrene.

\centering
\textbf{TOC Graphic}

\vspace{6pt}
\includegraphics[scale=1.0]{"figures/paper/main/TOC_graphic_v4.pdf"}

\end{abstract}


Despite significant advances, single-junction solar cells remain fundamentally constrained by the Shockley-Queisser limit.\cite{shockleyDetailedBalanceLimit1961} This barrier seriously hinders the transition toward multi-terawatt photovoltaics necessary to meet global energy demands.\cite{haegelPhotovoltaicsMultiterawattScale2023} One compelling strategy to surpass this limitation is spectral down-conversion via singlet fission (SF).\cite{tayebjeeShockleyqueisserMolecularApproaches2015, dayImprovingSpectralModification2019} SF occurs in organic materials and involves a spin-singlet ($\textrm{S} = 0$) exciton that converts into two lower-energy spin-triplet ($\textrm{S} = 1$) excitons, generally proceeding via an intermediate triplet-pair $^{1}\text{(TT)}$ state.\cite{Yago2016, collinsFluctuatingExchangeInteractions2019} If the independent triplet excitons can be successfully harvested, SF could be used to circumvent thermalization losses in photovoltaics. However, SF is highly sensitive to molecular structure and crystal packing,\cite{pensackExcitonDelocalizationDrives2015} requiring intensive synthetic and processing resources to optimize specific molecular geometries or crystalline architectures.

Modifying the photonic environment through strong light-matter coupling in microcavities may offer an alternative approach to control electronic and spin processes without structural alteration.\cite{martinez-martinezPolaritonAssistedSingletFission2018, climentNotDarkStrong2022, guOpticalCavityManipulationConical2021, zhangJointEffectsExcitonExciton2021} Strong light-matter coupling occurs when an ensemble of molecules interacts with a confined cavity photon mode. If the light-matter interaction rate exceeds the intrinsic decay rates of both the molecular excitations and the cavity photons, polariton splitting occurs and hybrid upper and lower polaritonic branches are formed.\cite{weisbuchObservationCoupledExcitonphoton1992, thompsonObservationNormalmodeSplitting1992, riderSomethingNothingLinking2021, Hopfield1958} These hybrid polaritonic states have been proposed to alter the photophysics of bare molecules, offering a potential method to control spin-dependent processes such as SF,\cite{martinez-martinezPolaritonAssistedSingletFission2018, climentNotDarkStrong2022, takahashiSingletFissionAmorphous2019, liuRoleLongLivedExcitons2020} triplet-triplet annihilation (TTA),\cite{polakManipulatingMoleculesStrong2020, siltanenEnhancingEfficiencyPolariton2025a} and intersystem crossing.\cite{eiznerInvertingSingletTriplet2019, siltanenEnhancingEfficiencyPolariton2025a}

However, manipulating chemical and spin dynamics via strong coupling remains controversial,\cite{imperatoreReproducibilityCavityenhancedChemical2021a, thomasSelectionBiasStrong2024, thomasNonPolaritonicEffectsCavityModified2024b} and experimental evaluations of polariton-mediated SF are limited. Prior investigations have focused on TIPS-pentacene\cite{liuRoleLongLivedExcitons2020} and rubrene\cite{takahashiSingletFissionAmorphous2019} strongly coupled within metallic silver microcavities.

For TIPS-pentacene, no modifications to SF kinetics were detected.\cite{liuRoleLongLivedExcitons2020} The authors suggested that a manifold of reservoir states dilutes the polariton dynamics. This exciton reservoir encompasses uncoupled molecules, higher-multiplicity spin states (such as triplets), and a degenerate manifold of dark states. Dark states arise from $N$ molecules interacting with a single photon mode, yielding two bright polariton states and $N-1$ dark states.\cite{herreraMolecularPolaritonsControlling2020a} In general, the exciton reservoir can populate the lower polariton branch (LPB) through different ``scattering" mechanisms,\cite{grantEfficientRadiativePumping2016, colesVibrationallyAssistedPolaritonRelaxation2011, litinskayaFastPolaritonRelaxation2004a} followed by rapid decay of the LPB to the ground state.\cite{wang_quantum_2014, michailAddressingDarkState2024} As the number of reservoir states far exceeds the number of polaritonic states in real systems, the reservoir typically dictates the dynamics.

Conversely, SF was reportedly suppressed in strongly coupled microcavities containing rubrene at large exciton-photon detunings, where the authors observed changes on picosecond timescales.\cite{takahashiSingletFissionAmorphous2019, Herrera2016} Reconciling these picosecond features with standard models of organic exciton-polaritons is challenging, however, given that the polariton lifetime is fundamentally bounded by the cavity lifetime, which typically remains below \SI{100}{fs} for lossy silver mirrors.\cite{hobsonStrongExcitonPhoton2002, monyAngleIndependentPolaritonEmission2018} The persistent picosecond signatures and reported suppression of SF in rubrene must therefore involve the exciton reservoir, motivating a deeper investigation into its role.

However, directly probing this reservoir with light remains difficult because the cavity acts as a narrow optical filter over the broad probe spectrum \cite{schwennickeWhenMolecularPolaritons2025}. In ultrafast transient absorption (TA) experiments, this filtering introduces pronounced pump-induced artifacts arising from ground-state bleaching (polariton contraction), thermal expansion, or refractive index changes,\cite{renkenUntargetedEffectsOrganic2021} which can severely complicate the analysis and introduce significant errors if not handled correctly.

To bypass these issues, we construct microcavities using dielectric distributed Bragg reflectors (DBRs) and track the excited-state dynamics of strongly coupled rubrene entirely within the near-infrared (NIR) region, outside the cavity stopband. By evaluating these systems across contrasting photon lifetimes and a range of exciton-photon detunings, we only find subtle cavity-induced features that do not reflect intrinsic modifications to SF. Instead, the exciton reservoir overwhelmingly dominates the photophysical response, leaving the SF dynamics virtually indistinguishable from uncoupled films. Our methods for this study can be found in the Supporting Information (SI), Section \ref{SI_sec_Experimental_methods}.

\begin{figure}[!tpb]
    \centering
    \includegraphics[width=0.8\linewidth]{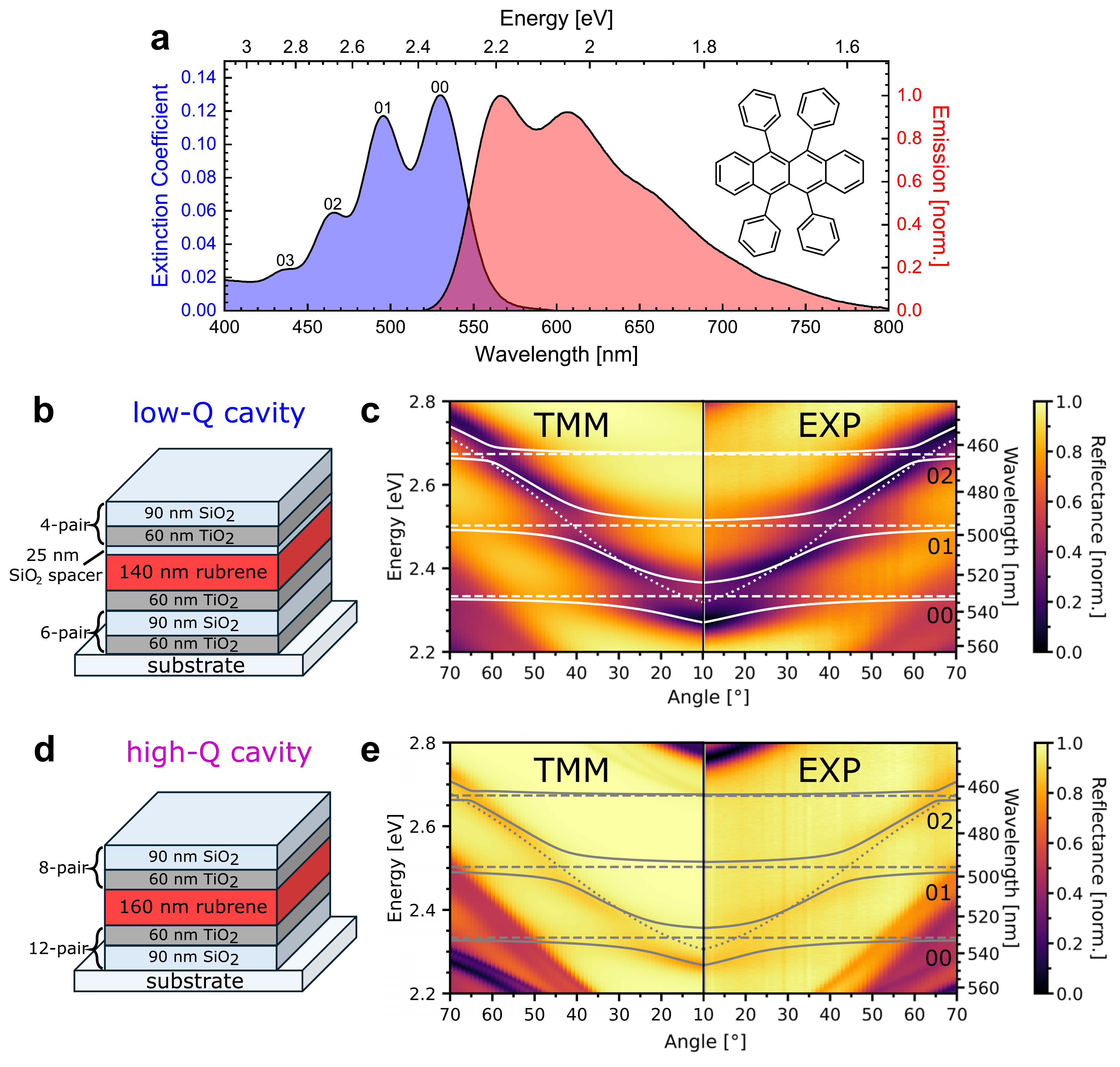}
       \caption{\textbf{Optical characterization of rubrene films and microcavities.} \textbf{(a)} Optical extinction coefficient (imaginary component of the complex refractive index, blue; see SI Subsection \ref{SI_sec_Experimental_methods_Steady_State_Optical_Characterization} for calculation) and emission (red) spectra of a \SI{140}{nm} rubrene film (inset: molecular structure) with labeled vibronic transitions. \textbf{(b, d)} Schematics of the \textbf{(b)} low-\(Q\) and \textbf{(d)} high-\(Q\) microcavities comprised of \(\text{SiO}_2/\text{TiO}_2\) distributed Bragg reflectors. \textbf{(c, e)} Experimental (EXP) and simulated (TMM) reflectance for the \textbf{(c)} low-\(Q\) and \textbf{(e)} high-\(Q\) cavities. Overlaid coupled-oscillator fits show cavity modes (dotted), vibronic transitions (dashed), and polariton branches (solid).}
    \label{fig_basic_characterisation}
\end{figure}

We begin by establishing the baseline photophysical properties of the active-layer rubrene thin films which have been prepared via thermal evaporation to enable direct comparison with previous experimental work.\cite{takahashiSingletFissionAmorphous2019} The extinction and emission spectra of a representative film are shown in Figure \ref{fig_basic_characterisation}a, with rubrene's molecular structure displayed in the inset. The extinction spectrum shows a clear vibronic progression, with energy spacings of $\sim \SI{0.17}{eV}$ between peaks corresponding to C--C stretching vibrational modes.\cite{leeLightEmittingRubreneNanowire2009, spanoSpectralSignaturesFrenkel2010} The intense 0-0 and 0-1 transitions maximize the light-matter coupling strength, while their relatively narrow linewidths enable clear resolution of the anti-crossings in reflectivity spectra (Figure \ref{fig_basic_characterisation}c,e).

As both strong coupling and SF can be impacted by the solid-state morphology, we characterize the film's architecture using Grazing Incidence Wide-Angle X-ray Scattering (GIWAXS, see SI Section \ref{SI_sec_GIWAXS}). The 2D GIWAXS map and its integrated intensity profiles reveal a weakly ordered morphology with a weak edge-on preference relative to the substrate surface. A similar GIWAXS pattern has been observed with ultra-smooth rubrene films,\cite{yoonUltrasmoothOrganicFilms2021} suggesting that our films are homogeneous, which is supported by our fluorescence lifetime imaging microscopy and stylus profilometry (see SI Section \ref{SI_sec_FLIM}).

We now turn to the cavity design and characterization. Considering ultra-fast processes such as coherent SF and near-instantaneous triplet formation are known to take place in rubrene on instrument-limited timescales (i.e. $< \SI{25}{\femto\second}$)\cite{beraFemtosecondRamanMicroscopy2017, breenTripletSeparationDrives2017, Miyata2017, maslennikovInterplayMixedPure2025, volekStructuralDisorderEdges2023, bossanyiSingletFissionIncoherent2024}, the polaritonic states may impact these processes directly if the cavity lifetime ($\tau_c$) is long enough.\cite{ostroverkhova_excitons_2024} We therefore fabricate two strongly coupled samples with contrasting quality ($Q$) factors: a low $Q$-factor microcavity ($Q=90$, $\tau_c \approx \SI{25}{\femto\second}$) and a relatively high $Q$-factor microcavity ($Q=1150$, $\tau_c \approx \SI{330}{\femto\second}$) by adjusting the number of dielectric pairs of the DBRs in each. Schematics of the microcavities are illustrated in Figure \ref{fig_basic_characterisation} and calculations of their Q-factors and lifetimes are detailed in SI Section \ref{SI_sec_cavity_mode_linewidth}. The DBR stopband is centered at \SI{530}{\nano\meter} to strongly couple rubrene's prominent ground-state absorption transitions, while simultaneously maintaining a transmissive NIR region to reduce artifacts in the NIR TA measurements. For an uncoupled control, we use an open-cavity (no top mirror) to fairly account for surface-born effects between rubrene and the bottom DBR---its structure and angular reflectance can be found in SI Section \ref{SI_sec_opencavity}.

The coupling strengths of the microcavities are determined using angle-resolved reflectance, which we present in Figure \ref{fig_basic_characterisation}. To quantify the Rabi splitting, a 4-level coupled oscillator model (solid lines) is manually fit to the measured dispersions. Here we use rubrene's strongest vibronic transitions described by Lorentzian oscillators (see SI Section \ref{SI_sec_strong_coupling_criterion} for extinction spectrum fitting) where we find an excellent match to the experimental data using Rabi splittings of approximately 100 meV for the 0-0 and 0-1 transitions, and 60 meV for the 0-2 transition. As the minimum Rabi splitting required for strong coupling for these systems is approximately 70-80 meV (calculation presented in SI Section \ref{SI_sec_strong_coupling_criterion}), the 0-0 and 0-1 transitions fall firmly in the strong coupling regime for both microcavities. The detunings (defined as the energetic separation between the 0-0 absorption transition and the cavity mode at normal incidence) also only deviate by $\sim\SI{10}{\milli\electronvolt}$ allowing us to solely compare the impact of their quality factors on SF.

\begin{figure}[!tbp]
    \centering
    \includegraphics[width=1\linewidth]{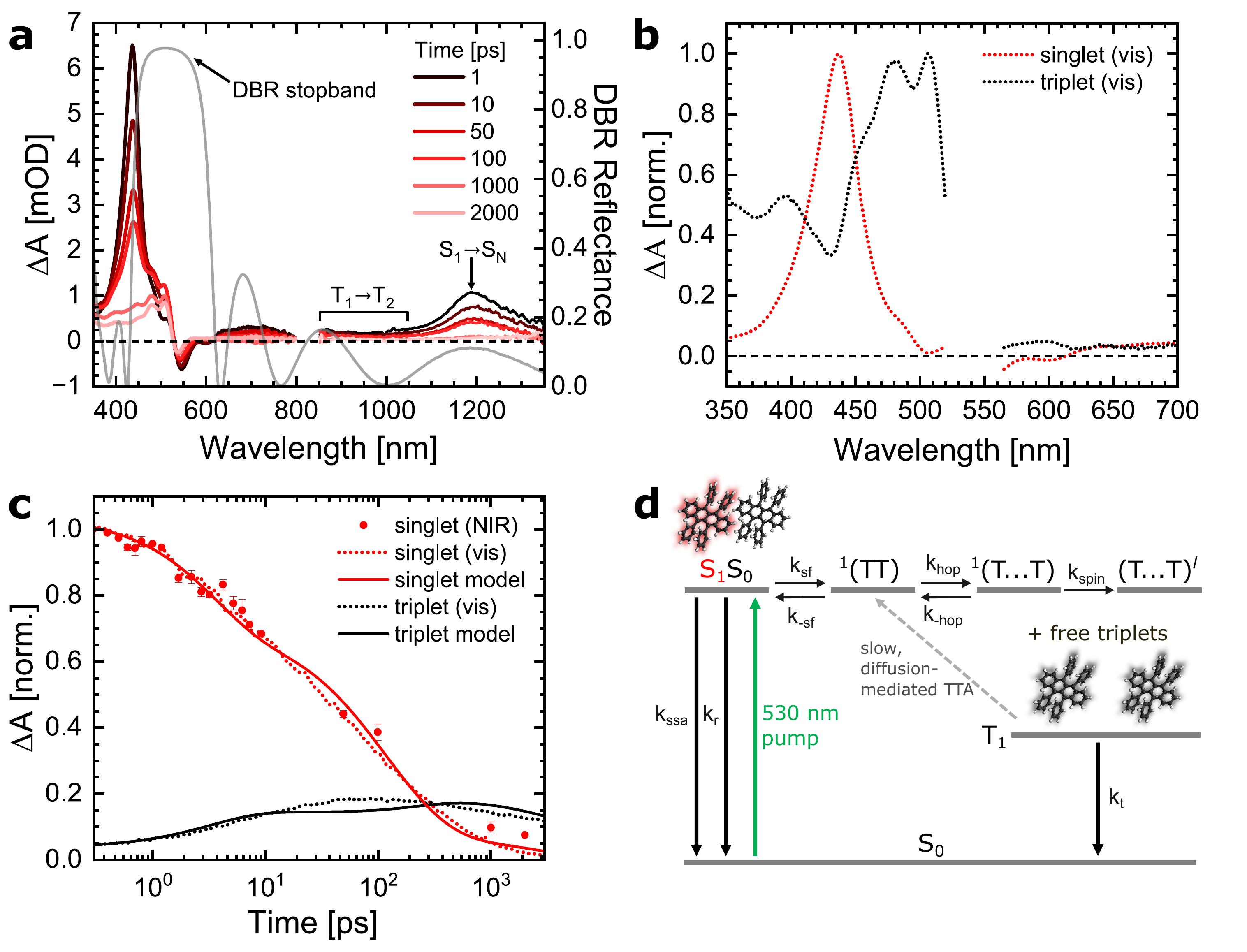}
       \caption{\textbf{Broad-band transient absorption of an uncoupled rubrene film.} \textbf{(a)} Time-dependent transient absorption ($\Delta A$) spectra spanning the visible (vis) and near-infrared (NIR) regions, following $\SI{530}{\nano\meter}$ excitation at $\SI{17}{\micro\joule\per\centi\meter\squared}$ fluence. Prominent NIR transitions are labeled ($\mathrm{T_{1} \to T_{2}}$ and $\mathrm{S_{1} \to S_{N}}$) and the reflectance of a 6-pair distributed Bragg reflector (DBR) is overlaid (gray line). \textbf{(b)} Rubrene's normalized singlet and triplet spectral components in the visible region extracted via Multi-Component Curve Resolution with Alternating Least Squares (MCR-ALS). \textbf{(c)} Normalized kinetic traces of the rubrene film associated with the visible (MCR-ALS) and NIR (average 1180--1220 nm, error bars are standard errors of 3 repeats) regions, overlaid with the singlet and triplet dynamics from the kinetic model. \textbf{(d)} Schematic outlining the kinetic model and associated rate constants describing rubrene's uncoupled photophysics.}
    \label{fig_SF_TA_kinetics}
\end{figure}

Next, to establish the baseline SF timescales of the uncoupled film and validate our NIR probing strategy, we assess its transient absorption signal across both the visible and NIR regions. Following excitation at \SI{530}{nm} rubrene undergoes efficient SF because its first singlet excited state ($\textrm{S}_1 \approx \SI{2.32}{eV}$) is roughly twice the energy of its triplet state ($\textrm{T}_1 \approx \SI{1.14}{eV}$).\cite{herkstroeterTripletStateEnergies1981, singh-rachfordPdIIPhthalocyanineSensitizedTripletTriplet2008} The corresponding broadband TA spectra are shown in Figure~\ref{fig_SF_TA_kinetics}a. Consistent with established literature,\cite{bossanyiSpinStatisticsTriplet2021} the early-time spectra are dominated by a strong singlet excited-state absorption (ESA) in the visible region at $\SI{440}{nm}$ and in the NIR region at $\SI{1200}{nm}$. Over time, a characteristic triplet ESA band emerges in the visible region at \SI{510}{nm},\cite{maSingletFissionRubrene2012a} while the mid-NIR window (850--1050 nm) contains overlapping contributions from both singlet and weaker $\textrm{T}_1 \to \textrm{T}_2$ transitions.\cite{bossanyiSpinStatisticsTriplet2021, Miyata2017}

We successfully isolate the singlet and triplet ESA contributions in the visible region via Multivariate Curve Resolution-Alternating Least Squares (MCR-ALS)\cite{jaumotMCRALSGUI202015}, and their extracted spectra are presented in Figure~\ref{fig_SF_TA_kinetics}b. The resulting time-resolved concentration profiles are then manually fit to a previously reported multi-step kinetic model,\cite{bossanyiOptimizedRubrenebasedNanoparticle2022} providing a consistent baseline to quantify kinetic variations between samples. The measured and simulated singlet kinetics of the bare film are overlaid and presented in Figure \ref{fig_SF_TA_kinetics}c, and a schematic of the kinetic model is presented in Figure \ref{fig_SF_TA_kinetics}d (see SI Section \ref{SI_sec_kinetic_model} for full rate modeling parameters). The model accounts for the initial interconversion to a spin-correlated triplet pair $^{1}\text{(TT)}$, subsequent spatial separation into independent pairs $^{1}\text{(T\dots T)}$, and long-term spin evolution into mixed states $\text{(T\dots T)}^l$.\cite{bossanyiOptimizedRubrenebasedNanoparticle2022, pilandMagneticFieldEffects2013b, wolfQuantumBeatsMultiexciton2018} Evaluating the residuals (the difference between the normalized experimental data and the kinetic simulations, see SI Section \ref{SI_constituent_singlet_dynamics}) reveals no systematic deviations, confirming a robust fit. The fit reveals that the primary $\textrm{S}_1 \to {^1\text{(TT)}}$ conversion occurs within \SI{8}{ps}, followed by spatial triplet separation with a time constant of $\sim\SI{125}{ps}$. These time constants are consistent with those previously reported for rubrene nanoparticles,\cite{bossanyiOptimizedRubrenebasedNanoparticle2022} with minor deviations expected due to differences in film morphology.

It is worth noting that the SF steps described by the kinetic model occur on timescales significantly longer than the photon lifetimes of our microcavities. Consequently, these processes are unlikely to be perturbed directly in the microcavities by the polaritonic modes prior to radiative photon leakage. Indeed, this was also the case with the previous SF report on strongly-coupled rubrene,\cite{takahashiSingletFissionAmorphous2019} although changes to SF on timescales exceeding the cavity lifetime were still observed. The changes reported may therefore stem from the exciton reservoir interacting with the polariton modes through energy scattering.\cite{georgiouGenerationAntiStokesFluorescence2018a, colesImagingPolaritonRelaxation2013} In our measurements, we also detect a clear non-zero triplet population within the instrument response function ($\sim\SI{100}{fs}$) which indicates coherent SF.\cite{beraFemtosecondRamanMicroscopy2017, breenTripletSeparationDrives2017, Miyata2017, maslennikovInterplayMixedPure2025, volekStructuralDisorderEdges2023, bossanyiSingletFissionIncoherent2024} This ultra-fast channel is shorter than the photon lifetime of the high-$Q$ microcavity ($\tau_c \approx \SI{330}{\femto\second}$), making it a prime candidate for direct polaritonic modulation.

To probe the microcavities successfully we must bypass the extensive pump-induced optical artifacts associated with their confinement in the visible domain.\cite{renkenUntargetedEffectsOrganic2021} As we show in Figure~\ref{fig_SF_TA_kinetics}a, the DBR mirror reflectance drops close to zero in the NIR region, minimizing the risk of artifacts here. We therefore probe the microcavity dynamics in the NIR region. To validate the NIR singlet peak as a proxy for the visible domain, we overlay the $\sim\SI{1200}{nm}$ kinetic trace with the visible region's singlet kinetics in Figure~\ref{fig_SF_TA_kinetics}c. The strong agreement between visible and NIR singlet dynamics confirms that the NIR singlet peak serves as an unambiguous proxy for the underlying SF dynamics.

With the NIR probe validated, we now compare the TA signal of the uncoupled open cavity to those of the strongly coupled microcavities. We again pump the samples at $\SI{530}{\nano\meter}$ to directly excite the exciton reservoir. We have carefully controlled the pump power to minimise singlet-singlet exciton annihilation, a population-dependent process that occurs on picosecond timescales which can produce misleading kinetics and spectra if not effectively managed.\cite{bossanyiOptimizedRubrenebasedNanoparticle2022, leSlowSingletFission2016} This was achieved by using low pump pulse energies to account for differences in absorption between samples by approximately matching the initial intensity of the $\mathrm{S_{1} \to S_{N}}$ ($\SI{1200}{\nano\meter}$) transition. As a consequence, the signal-to-noise ratio is lower than typical TA measurements, though the $\SI{1200}{\nano\meter}$ peak is still clearly resolved.

\begin{figure}[!tbp]
    \centering
    \includegraphics[width=0.47\linewidth]{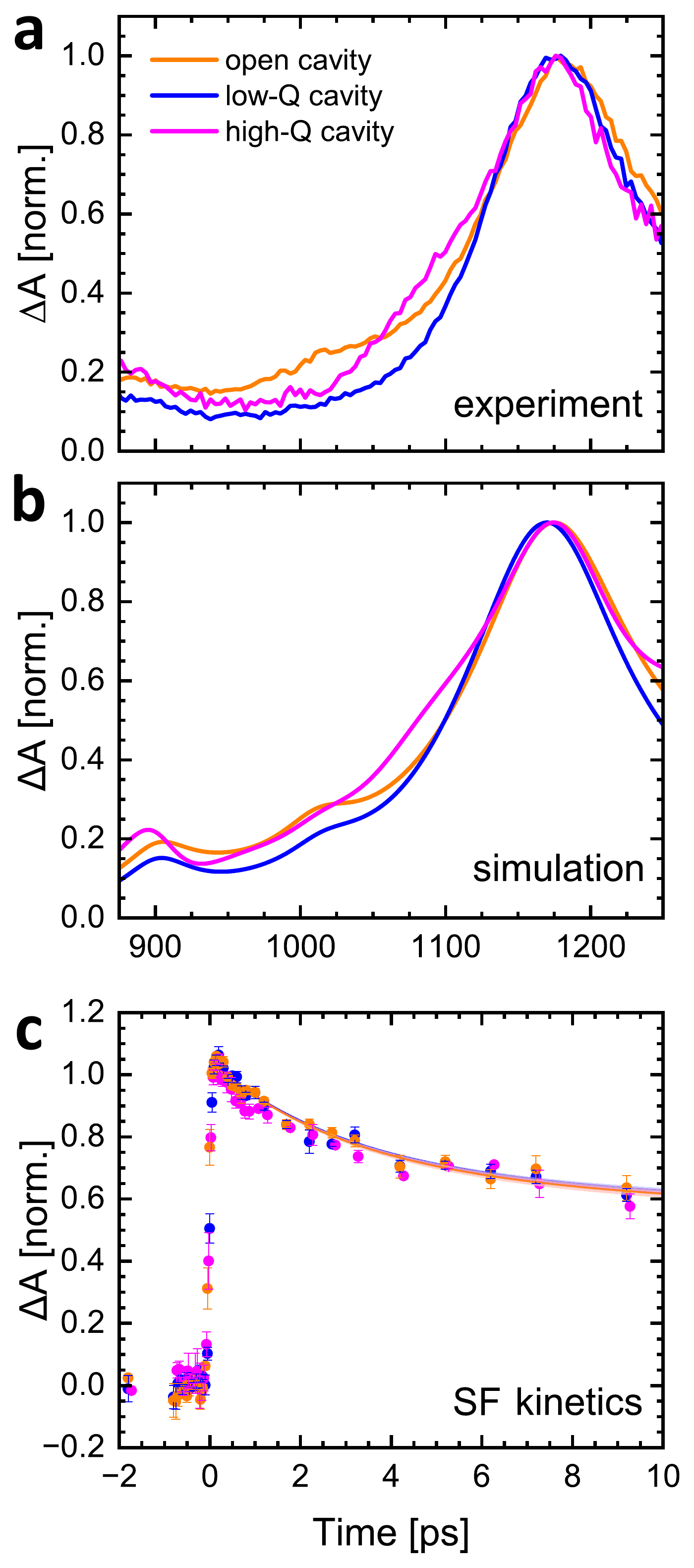} \caption{\textbf{Near-infrared transient absorption of coupled microcavities and uncoupled controls.} \textbf{(a)} Prompt transient absorption spectra of the low-$Q$/high-$Q$ microcavities and open cavity control, averaged between 0.1-0.3 ps and normalized to the maximum signal. \textbf{(b)} Excited-state transfer matrix spectral simulation assuming $\SI{0.2}{\percent}$ ground-state bleaching upon photo-excitation. \textbf{(c)} Normalized singlet (1180--1220 nm) dynamics overlaid with the kinetic model (solid line). The kinetics are taken from the average of three low-fluence measurements (see SI Section \ref{SI_constituent_singlet_dynamics}) normalized to the average signal between 0.2--0.5 ps; error bars are the standard error of the mean normalized kinetic.}
    \label{fig_cavity_TA_loqQ_highQ}
\end{figure}

In Figure \ref{fig_cavity_TA_loqQ_highQ}a we present the normalized spectra averaged over the early-time window ($\SI{0.1}{\pico\second}\text{--}\SI{0.3}{\pico\second}$). This prompt measurement is taken within the photon lifetime of the high-$\mathrm{Q}$ cavity, allowing us to assess whether the cavity's coherence has an effect on $\mathrm{SF}$. The full time-resolved spectra for each sample is presented in SI Section \ref{SI_sec_cavity_spectra}. 

With both microcavities, we find subtle spectral changes relative to both the open cavity and bare film controls, which could indicate modification of coherent SF.\cite{maslennikovInterplayMixedPure2025} To understand these changes, we have conducted thorough transfer matrix method (TMM) simulations of the TA spectra; further details can be found in SI Section \ref{SI_TA_TMM}. In Figure \ref{fig_cavity_TA_loqQ_highQ}b we present a simulation of the TA spectra where the transmission of the samples is calculated in the ground-state (pump-off) and excited-state (pump-on) configurations, before calculating the TA spectrum through $\Delta A = - \log_{10} \left( T_{\text{on}}/T_{\text{off}} \right)$. Importantly, the simulation reproduces the majority of the spectral changes we have measured, successfully capturing cavity spectra even at wide probing angles (see SI Figure \ref{fig_cavity_TA_spectra_angular_modelling}). This indicates that the subtle changes to the prompt spectra predominantly stem from optical effects associated with the mirrors rather than changes to coherent SF, despite the high-$\mathrm{Q}$ cavity's prolonged photon lifetime.

The simulation can be optimized to account for pump-induced changes such as ground-state bleaching (polariton-contraction), thermal expansion,\cite{shengActiveControlElectromagnetic2015} or changes to refractive index, which has been explored previously in the visible domain.\cite{renkenUntargetedEffectsOrganic2021} In SI Section \ref{SI_TA_TMM} we present a comprehensive evaluation of these effects on our samples across the visible and NIR regions using the TMM approach. Although we do find that these pump-induced spectral artifacts impact the spectra in the NIR, the changes are subtle and far less pronounced than in the visible region. We can therefore accurately extract the NIR singlet dynamics without risk of them being contaminated by severe untargeted effects\cite{renkenUntargetedEffectsOrganic2021}.

The singlet dynamics in Figure \ref{fig_cavity_TA_loqQ_highQ}c are very similar for both the strongly coupled microcavities and open cavity control. As such, it is unnecessary to develop a new kinetic model to describe the singlet dynamics of the strongly coupled systems. Instead, we apply the model outlined in Figure \ref{fig_SF_TA_kinetics}d across all architectures; the optimized rate constants are listed in SI Section \ref{SI_sec_kinetic_model}. The full dynamics, encompassing the kinetic steps following the initial SF phase, are also presented in SI Section \ref{SI_constituent_singlet_dynamics}.

Crucially, the kinetics of the uncoupled open cavity and strongly coupled microcavities can be described by identical model parameters across all measured timescales. For each architecture, the initial SF step, $\text{S}_1 \to {^1\text{(TT)}}$, proceeds with an optimized rate constant of $k_{\text{SF}} = 130\text{ ns}^{-1}$. The residuals of the normalized kinetics (SI Section \ref{SI_constituent_singlet_dynamics}) are similar across the sample set, with no systematic deviations, indicating a good fit. To quantify the fit, we calculated the root mean square error (RMSE) of each sample's residuals, yielding identical values of $\text{RMSE} = 0.038$ for both the open cavity and low-$\mathrm{Q}$ cavity, and $\text{RMSE} = 0.061$ for the high-$\mathrm{Q}$ cavity. The mildly elevated error for the high-$\mathrm{Q}$ cavity is likely due to a lower signal-to-noise ratio (from reduced molecular absorption) and increased susceptibility to cavity-induced artifacts (see SI Figure \ref{SI_fig_sim_TA_artefacts_highQ_cavity}). Nevertheless, the experimental kinetics presented in Figure \ref{fig_cavity_TA_loqQ_highQ}c show that the high-$\mathrm{Q}$ cavity has extremely similar dynamics to both the low-$\mathrm{Q}$ cavity and the uncoupled open cavity. This agreement across the three architectures effectively rules out the possibility that strong coupling alone perturbs SF in these rubrene samples.

\begin{figure}
    \centering
    \includegraphics[width=0.47\linewidth]{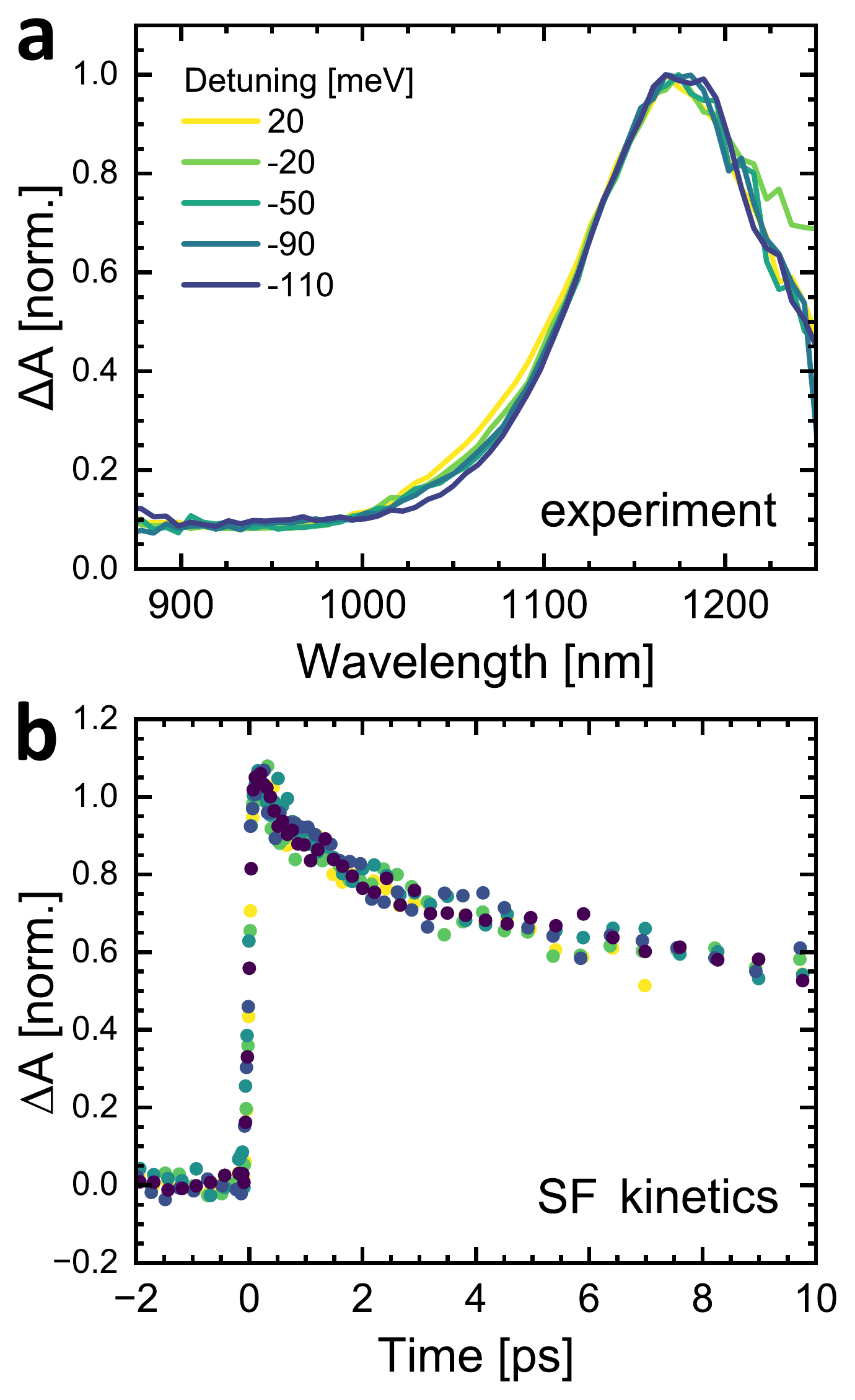} \caption{\textbf{Near-infrared transient absorption microcavity detuning dependence.} \textbf{(a)} $\Delta A$ signal for a range of exciton-cavity photon detunings, averaged between 0.1-0.3 ps and normalized to the maximum signal. The pump fluences from positive to negative detuning ($\SI{20}{meV}$ to $\SI{-110}{meV}$) are $\SI{10}{\micro\joule\per\centi\meter\squared}$, $\SI{10}{\micro\joule\per\centi\meter\squared}$, $\SI{10}{\micro\joule\per\centi\meter\squared}$, $\SI{15}{\micro\joule\per\centi\meter\squared}$ and $\SI{32}{\micro\joule\per\centi\meter\squared}$ to maintain a similar initial peak $\Delta A$ signal ($\sim \SI{1.6}{mOD}$) and control singlet-singlet exciton annihilation. \textbf{(b)} Cavity singlet (1180--1220 nm) dynamics normalized to the average initial peak signal between 0.2--0.5 ps.}
    \label{fig_cavity_TA}
\end{figure}

However, given that rubrene's SF dynamics are reportedly modulated by exciton-photon detuning in the strong coupling regime,\cite{takahashiSingletFissionAmorphous2019} we have fabricated microcavities with a range of detunings to test this. The detuned structures are similar to the low-$Q$ microcavity but with varying $\text{SiO}_2$ spacer thicknesses (between $\SI{0}{\nano\meter}$ and $\SI{50}{\nano\meter}$) to tune the cavity thickness and thus optical resonance. Their fitted dispersions are presented in SI Section \ref{SI_sec_detunings}. We examine a detuning range from $\SI{20}{\milli\electronvolt}$ to $\SI{-110}{\milli\electronvolt}$, which encompasses and extends past the $\SI{-80}{\milli\electronvolt}$ detuning where SF was reportedly modified.\cite{takahashiSingletFissionAmorphous2019} In the prompt spectra (Figure \ref{fig_cavity_TA}a) we again see no major spectral changes, which suggests coherent SF is not perturbed by detuning. There are also no significant changes to the kinetics (Figure \ref{fig_cavity_TA}b). We therefore find no evidence of SF being perturbed by detuning or cavity lifetime within the picosecond time-scales explored in these TA measurements.

Finally, our NIR TA results are further supported by transient electron paramagnetic resonance (EPR) measurements, which inherently bypass optical cavity artifacts by probing spin dynamics via magnetic dipole transitions rather than optical transmission. Transient EPR can probe the nature and dynamics of high-spin states, such as free triplets, formed by SF on later timescales. As spin must be conserved in SF, the process initially proceeds via a spin-correlated triplet pair of singlet character, $^{1}\text{(TT)}$. Singlet–quintet spin mixing can then occur in the weak inter-triplet exchange-coupling regime, prior to spin decoherence and the formation of free triplets.\cite{taitComputationalToolsSimulation2023, musserTripletPairStatesOrganic2019} Since the spin-correlated triplet pair initially has singlet character, the $m_S = 0$ sublevels of the quintet and free triplet states are preferentially populated,\cite{tayebjeeQuintetMultiexcitonDynamics2017, weissStronglyExchangecoupledTriplet2017} leading to a non-equilibrium sublevel population distribution and a distinct spectral polarization pattern associated with SF.\cite{taitComputationalToolsSimulation2023} However, upon comparing the uncoupled rubrene film and low-$\mathrm{Q}$ cavity (see SI Section \ref{SI_sec_EPR}), there are no obvious differences; the spectral features remain virtually identical in shape and decay. This indicates that the dynamics of these triplet states and their spin-dependent formation are not perturbed by strong coupling, aligning with our TA results.

In summary, we have demonstrated that probing dielectric microcavities outside their reflective stopband using TA is an effective method for inspecting the photophysics of exciton-polariton systems that avoids optical artifacts. Using this method, we found no changes in the SF dynamics associated with strong light-matter coupling in rubrene, even when the detuning and quality factor are varied. Although minor spectral deviations were observed, transfer matrix modelling successfully accounts for these features, showing that they originate primarily from interference effects associated with the mirrors rather than changes to molecular photophysics. This result casts significant doubt on previous reports of polariton-modulated SF in rubrene\cite{takahashiSingletFissionAmorphous2019} but agrees with the conclusions reported for strongly coupled TIPS-pentacene that the exciton reservoir dominates the dynamics.\cite{liuRoleLongLivedExcitons2020} We suggest that the previously reported changes with rubrene likely stem from optical artifacts, surface effects, or second-order dynamics, all of which were thoroughly controlled in this work. More broadly, these untargeted effects must be considered in other studies involving microcavities before assigning changes to strong coupling.

Since the presence of the exciton reservoir appears to negate any perturbation of SF, it must be avoided in future investigations. This could be achieved by using higher coupling strengths that spectrally isolate the LPB to reduce back-transfer to the reservoir. As previously theorized,\cite{climentNotDarkStrong2022, martinez-martinezPolaritonAssistedSingletFission2018} this approach would be applicable to highly exothermic SF materials where the LPB and the triplet-pair state can be brought into resonance through strong coupling, but it would require selectively exciting the LPB which may limit its practical use in spectral conversion. A more functional avenue for cavity-modified spin processes may then be through TTA, which already shows promising evidence.\cite{berghuisEnhancedDelayedFluorescence2019, polakManipulatingMoleculesStrong2020} Crucially, to unravel these modified photophysics, our results demonstrate that probing outside the stopband of dielectric mirrors provides a simple yet effective framework to isolate genuine transient absorption signals that can be applied to future microcavity experiments.

\begin{acknowledgement}

J.C., M.R., R.C.K., R.H.G., J.P.P., G.J.L., and T.Y. thank UKRI (EPSRC grant codes EP/T012455/1, EP/X525790/1, and EP/Y031962/1) and the University of Sheffield for support. C.E.T. acknowledges the Royal Society for a University Research Fellowship (URF\textbackslash R1\textbackslash 201071).

Funding from the EPSRC for the Lord Porter Laser Laboratory (EP\slash R042802\slash 1, EP\slash L022613\slash 1) and for the Xenocs\slash Excillum X-ray scattering instrument (EP\slash M028437\slash 1, EP\slash V034804\slash 1) is gratefully acknowledged.

The authors thank Ravi Kumar Venkatraman for assistance with preliminary experiments and early-stage data collection, and Rahul Jayaprakash for providing the transfer matrix modeling code used to simulate and design the microcavity architectures.







\end{acknowledgement}


\section*{Data Availability}
Data will be available on the University of Sheffield’s repository, ORDA.

\begin{suppinfo}


The Supporting Information is available free of charge on the ACS Publications website at DOI: [Insert DOI].

Experimental methods; GIWAXS morphology characterization; fluorescence lifetime imaging microscopy decay maps; quality factors and strong coupling threshold calculations; excited-state TMM modeling; kinetic rate modeling of excited-state dynamic; and transient EPR data (Insert PDF download link).

\end{suppinfo}


\clearpage

\renewcommand{\thepage}{S\arabic{page}}
\setcounter{page}{1}
\renewcommand{\thesection}{} 
\setcounter{section}{1}
\renewcommand{\thesubsection}{S\arabic{subsection}}
\renewcommand{\thetable}{S\arabic{table}}
\setcounter{table}{1}
\renewcommand{\thefigure}{S\arabic{figure}}
\setcounter{figure}{1}
\renewcommand{\theequation}{S\arabic{equation}}
\setcounter{equation}{1}


\section*{Supporting Information (SI)}

\subsection{Experimental Methods}\label{SI_sec_Experimental_methods}

\subsubsection{Rubrene Thin Film Fabrication} All materials were deposited onto clean quartz-coated glass (S151, Ossila) substrates following sequential sonication for \SI{10}{min} each in a Hellmanex~III/deionized (DI) water solution, pure DI water, and isopropyl alcohol. The substrates were then dried under a continuous nitrogen stream and subjected to a \SI{10}{min} UV-ozone treatment. Rubrene thin films ($>99\%$, Sigma Aldrich) were subsequently deposited via thermal evaporation (Angstrom Engineering) under high vacuum ($\leq\SI{2e-6}{mbar}$) at a sublimation temperature of approximately \SI{190}{\degree C}. A constant deposition rate of \SI{0.2}{nm.s^{-1}} was maintained and monitored using a quartz crystal microbalance, calibrated via a tooling factor verified by a Bruker DektakXT stylus profilometer. To prevent photo-oxidation, all bare films and open cavity controls were immediately encapsulated with a \SI{150}{nm} layer of $\text{SiO}_2$ deposited via electron-beam evaporation.

\subsubsection{Steady-State Optical Characterization}\label{SI_sec_Experimental_methods_Steady_State_Optical_Characterization} Steady-state transmittance spectra were acquired using an Agilent Cary 60 UV-vis spectrophotometer. Baseline corrections were performed using bare substrates to account for substrate reflections, and dark-current sweeps were subtracted to eliminate electronic instrument noise.

To determine the fundamental optical properties of the bare rubrene films, the dimensionless optical extinction coefficient ($k$) spectrum was derived directly from the baseline-corrected transmittance ($T$, expressed as a fraction) across the wavelength ($\lambda$) range according to the relationship:
\begin{equation}
k(\lambda) = \frac{-\ln[T(\lambda)] \cdot \lambda}{4\pi \cdot d}
\end{equation}
where $d$ is the thickness of the film.

Angle-resolved reflectance maps of the microcavities were collected from \ang{10} to \ang{70} in \ang{1} steps using a custom-built goniometer setup. Unpolarised white light from a deuterium–tungsten lamp (Ocean Optics DH-2000-BAL) was collimated and focused onto the sample at the goniometer center. The specularly reflected light was collected via a symmetric detection arm, fiber-coupled to an Andor Shamrock SR-303i-A triple-grating CCD spectrometer, and calibrated against a planar silver reference mirror to establish absolute reflectance before normalizing to the maximum signal.

Steady-state photoluminescence (emission) of the rubrene film was recorded using the same custom-built goniometer setup. A continuous-wave \SI{405}{nm} diode laser (Thorlabs) was focused onto the sample and emission collected at normal incidence (\ang{0}) relative to the sample surface. A long-pass filter ($\lambda > \SI{450}{nm}$) was positioned immediately before the spectrometer to reject the excitation beam.

\subsubsection{Microcavity Design and Fabrication} Planar microcavities were designed using a custom transfer matrix method (TMM) model to simulate multi-interfacial reflections and optimize layer thicknesses. The complex refractive indices ($\tilde{N} = n + ik$) of the organic active layers were determined by mapping the experimental extinction coefficient ($k$) with Lorentzian oscillators, with the dispersive real component ($n$) calculated via a Kramers-Kronig transform. Optical constants for the dielectric mirror layers were verified via spectroscopic ellipsometry (J.A. Woollam M-2000) at an incidence angle of \SI{75}{\degree}. Distributed Bragg reflectors (DBRs) were fabricated by depositing alternating quarter-wave ($\lambda/4$) layers of $\text{SiO}_2$ and $\text{TiO}_2$ ($>99.9\%$, Kurt J. Lesker) under high vacuum ($\leq\SI{5e-6}{mbar}$) using electron-beam evaporation (Angstrom Engineering). Deposition rates and target thicknesses were monitored via a quartz crystal microbalance.

\subsubsection{Grazing-Incidence Wide-Angle X-ray Scattering (GIWAXS)} Structural characterization was performed using a Xeuss 2.0 laboratory beamline (Xenocs) equipped with a liquid gallium MetalJet (Excillum) source operating at an energy of \SI{9.243}{keV} ($\lambda = \SI{1.34}{\angstrom}$). Samples were deposited onto Si(100) substrates to eliminate amorphous background scattering from glass. The collimated beam was incident on the sample at a fixed grazing angle of \SI{0.15}{\degree}. The scattered X-rays were collected under vacuum to minimise air scattering, using a Pilatus3R 1M 2D detector (Dectris) placed at a calibrated sample-to-detector distance of approximately \SI{0.31}{m}. The detector distance was verified using a silver behenate standard in transmission geometry. The resulting 2D data frames were geometrically corrected, masked, and converted into reciprocal space coordinates ($q$) using the pyFAI and pygix Python libraries.\cite{ashiotisFastAzimuthalIntegration2015}

\subsubsection{Fluorescence Lifetime Imaging Microscopy (FLIM)} Time-resolved fluorescence maps were acquired using a custom-built microscope setup. Samples were excited at a \SI{1}{MHz} repetition rate using a \SI{485}{nm} pulsed laser diode (PicoQuant, PDL828 laser driver) focused through a $20$$\times/$$0.5$ air objective (Zeiss, EC Epiplan-Neofluar). The collected emission was isolated using a \SI{550}{nm} long-pass filter to reject scattered excitation light, and single-photon arrival times were recorded using a time-correlated single-photon counting (TCSPC) module (Becker \& Hickl, SPC-150) synchronised to the laser trigger. The resulting fluorescence lifetime decay maps were processed and analyzed using the FLIMfit 5.1.1 software package.

\subsubsection{Transient Absorption (TA) Spectroscopy} Ultrafast TA measurements were performed using a commercial spectrometer (Helios, Ultrafast Systems) powered by a amplified Ti:Sapphire laser system (Spectra-Physics MaiTai seed and Spitfire Ace PA-40 chirped-pulse amplifier). The amplifier outputted \SI{800}{nm} pulses (\SI{40}{fs}, \SI{10}{kHz}, \SI{12}{W}). Excitation pulses at \SI{530}{nm} were generated using an optical parametric amplifier (TOPAS Prime, Light Conversion) coupled with a frequency mixer (NirUVis, Light Conversion), with the pump fluence adjusted via a neutral density filter prior to an optical chopper that modulated the pump frequency to \SI{5}{kHz}. The broadband probe was generated by focusing the \SI{800}{nm} fundamental into various non-linear media: a continuously translating $\text{CaF}_2$ crystal for the UV-vis region (360--780 nm), or a YAG crystal for the NIR (820--1350 nm). The relative pump–probe time delay was controlled via a motorized linear delay stage operating with a randomized spatial stepping order. Pump and probe polarizations were set at the magic angle (\ang{54.7}), and transmitted probe intensities were recorded using an InGaAs array detector. Raw datasets were background-subtracted and numerically corrected for group velocity dispersion (spectral chirp). Overlapping singlet and triplet excited-state spectral signatures in the visible were deconvoluted via multivariate curve resolution alternating least squares (MCR-ALS) using sub-picosecond and post-nanosecond data profiles, respectively, as initial spectral constraints.

\subsubsection{Transient Electron Paramagnetic Resonance (trEPR)} To probe triplet state dynamics on the nanosecond-to-microsecond timescale, trEPR measurements were conducted at the University of Oxford following a protocol optimized for organic thin films.\cite{bualiIntersystemCrossingOutcompetes2025} Samples were broken into fragments and loaded into high-purity quartz EPR tubes. Spectra were recorded on a Bruker Elexsys E680 X-band spectrometer equipped with a Bruker ER-4118X-MD5-W1 dielectric resonator. Cryogenic environments were maintained via an Oxford Instruments CF935 helium cryostat. Photo-excitation at \SI{505}{nm} was provided by a diode-pumped Q-switched Nd:YAG laser with an integrated optical parametric oscillator (EKSPLA NT230), operating at a \SI{50}{Hz} repetition rate with a \SI{5}{ns} pulse width. Isotropic photo-excitation was achieved by placing an achromatic depolarizer after the final turning mirror, and laser pulse energy at the sample was calibrated with a thermal detector (Thorlabs TD2X). Synchronization between the pulsed laser and the microwave bridge was managed via a digital delay generator (Stanford Research Systems DG645). Direct-detection trEPR transients were recorded without lock-in modulation using a transient recorder (Video Amplifier III, \SI{1}{GHz} bandwidth) with a continuous microwave excitation at a power of \SI{2}{mW}. Background laser artifacts were numerically subtracted using a two-dimensional baseline-correction matrix derived from off-resonance transients collected at both high and low magnetic fields.

\subsection{Grazing-incidence wide-angle X-ray scattering}\label{SI_sec_GIWAXS}

\begin{figure}[H]
    \centering
    \includegraphics[width=0.8\linewidth]{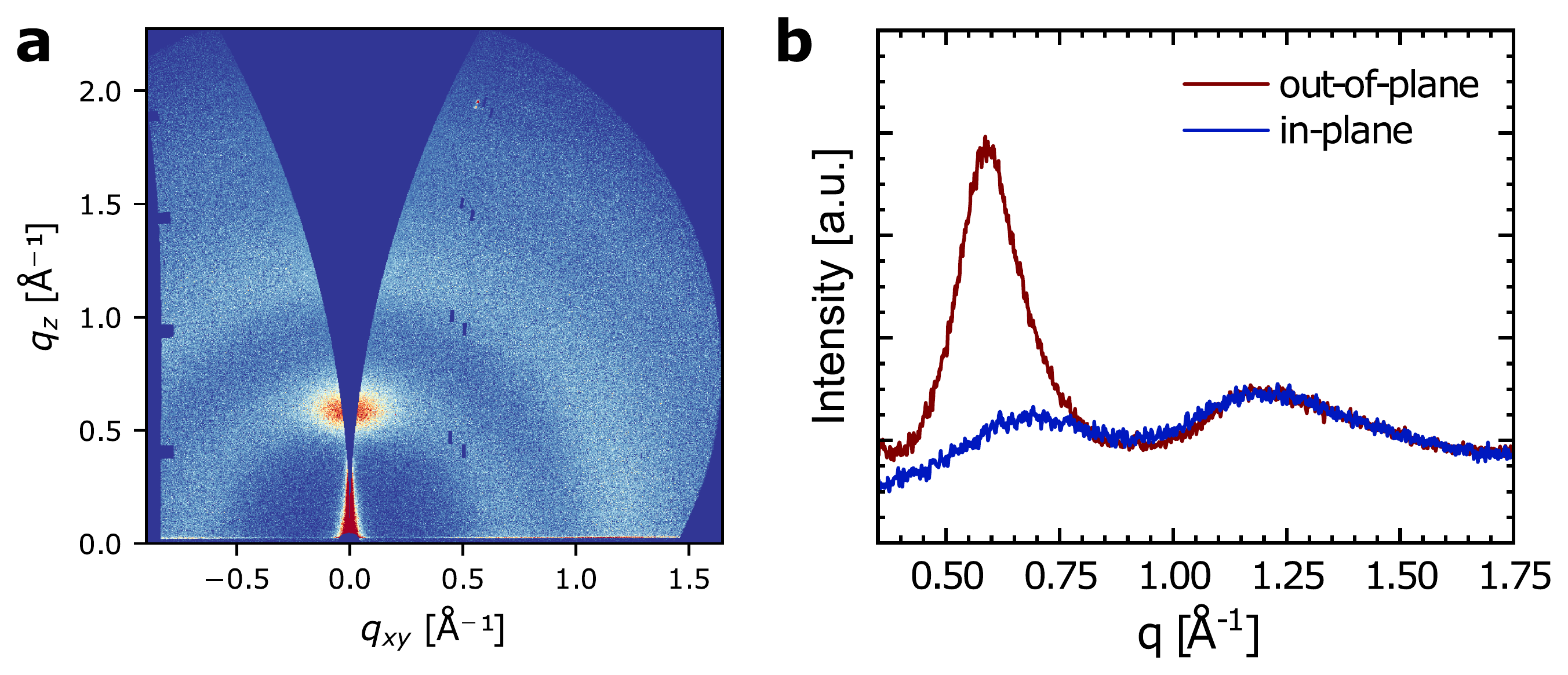}
    \caption{\textbf{Grazing-incidence wide-angle X-ray scattering (GIWAXS) characterization of a thermally evaporated rubrene film} \textbf{(a)} GIWAXS pattern of a rubrene film prepared by thermal evaporation indicating a weakly ordered structure. \textbf{(b)} Azimuthally integrated $q$-dependent 1D intensity profiles of the GIWAXS data integrated through various $\chi$ angles normal to the beam incidence at the detector: out-of-plane (in the $q_z$ direction, $\chi = 0^\circ \pm 20^\circ$) and in-plane (which includes all other angles $20^\circ \leq \chi \leq 90^\circ$), indicating an out-of-plane stacking preference.}
    \label{SI_fig_GIWAXS}
\end{figure}

The 2D GIWAXS map and its integrated intensity profiles, shown in Figure \ref{SI_fig_GIWAXS}, reveal a weakly ordered morphology, indicated by the diffuse, low-intensity, isotropic rings. This suggests that some weak short-range order is present in the film but limited long-range crystalline order. However, a closer analysis of the intensity profiles (Figure \ref{SI_fig_GIWAXS}b) shows a clear preference in the out-of-plane ($q_z$) direction at approximately $q_1 \approx \SI{0.6}{\text{\AA}^{-1}}$, with a less intense peak observed at $q_2 \approx \SI{1.2}{\text{\AA}^{-1}}$ that has no significant orientational preference. The peak at $\SI{0.6}{\text{\AA}^{-1}}$ corresponds to a lattice spacing of $d = 2\pi/q_1 \approx \SI{10.5}{\text{\AA}}$, which is similar to known lattice parameters for rubrene crystal polymorphs (e.g., the $c$-axis of the triclinic phase is $\approx \SI{12}{\text{\AA}}$)\cite{bergantinPressureInducedConformationalChange2014} and is characteristic of the distance between stacked molecular units. The second peak at $\SI{1.2}{\text{\AA}^{-1}}$ corresponds to a lattice spacing of $d = 2\pi/q_2 \approx \SI{5.2}{\text{\AA}}$. The out-of-plane preference of this scattering feature suggests that the rubrene molecules predominantly adopt a weakly ordered arrangement with weak edge-on preferential orientation relative to the substrate surface.

\subsection{Fluorescence life-time imaging microscopy}\label{SI_sec_FLIM}

\begin{figure}[H]
    \centering
    \includegraphics[width=0.8\linewidth]{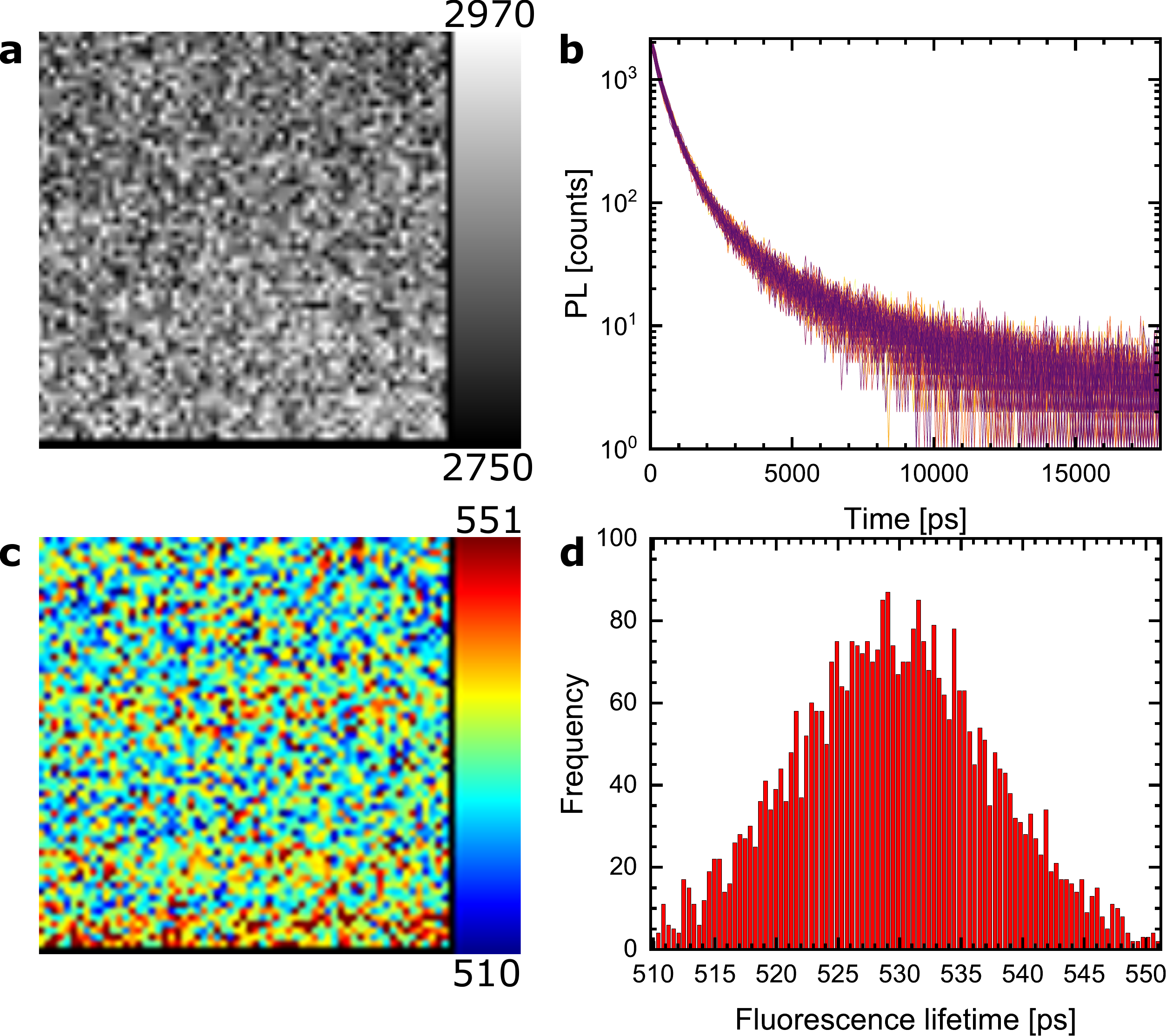}
    \caption{\textbf{Fluorescence lifetime imaging microscopy (FLIM) of a polycrystalline rubrene thin film.} \textbf{(a)} Integrated photon count map ($\SI{16 x 16}{\micro\meter}$), showing the fluorescence intensity detected immediately following excitation. \textbf{(b)} Fluorescence decay kinetics of 100 randomly selected points from the map, showing the signal decay over time. \textbf{(c)} Fluorescence lifetime map ($\SI{16 x 16}{\micro\meter}$), where the fluorescence lifetime is determined by fitting the instrument response corrected decay curve for each pixel with a mono-exponential decay. \textbf{(d)} Histogram of the fluorescence lifetime distribution corresponding to the fluorescence lifetime map in panel c.}
    \label{SI_fig_FLIM}
\end{figure}

There is no clear relationship between the fluorescence lifetime or intensity, and the position on the map, suggesting that the films are homogeneous on this scale (\SI{250}{\nano\meter} pixel size). The intensity and lifetime range is also narrow (given by the minimum and maximum values on the colour bar), confirming the films are a similar thickness across the maps area. Stylus profilometry thickness measurements (not shown) also exhibit minimal variance (standard deviation $< \SI{5}{\nano\meter}$) across the sample, further evidencing homogeneous films.

\subsection{Cavity lifetime calculation}
\label{SI_sec_cavity_mode_linewidth}

\begin{figure}[H]
    \centering
    \includegraphics[width=0.6\linewidth]{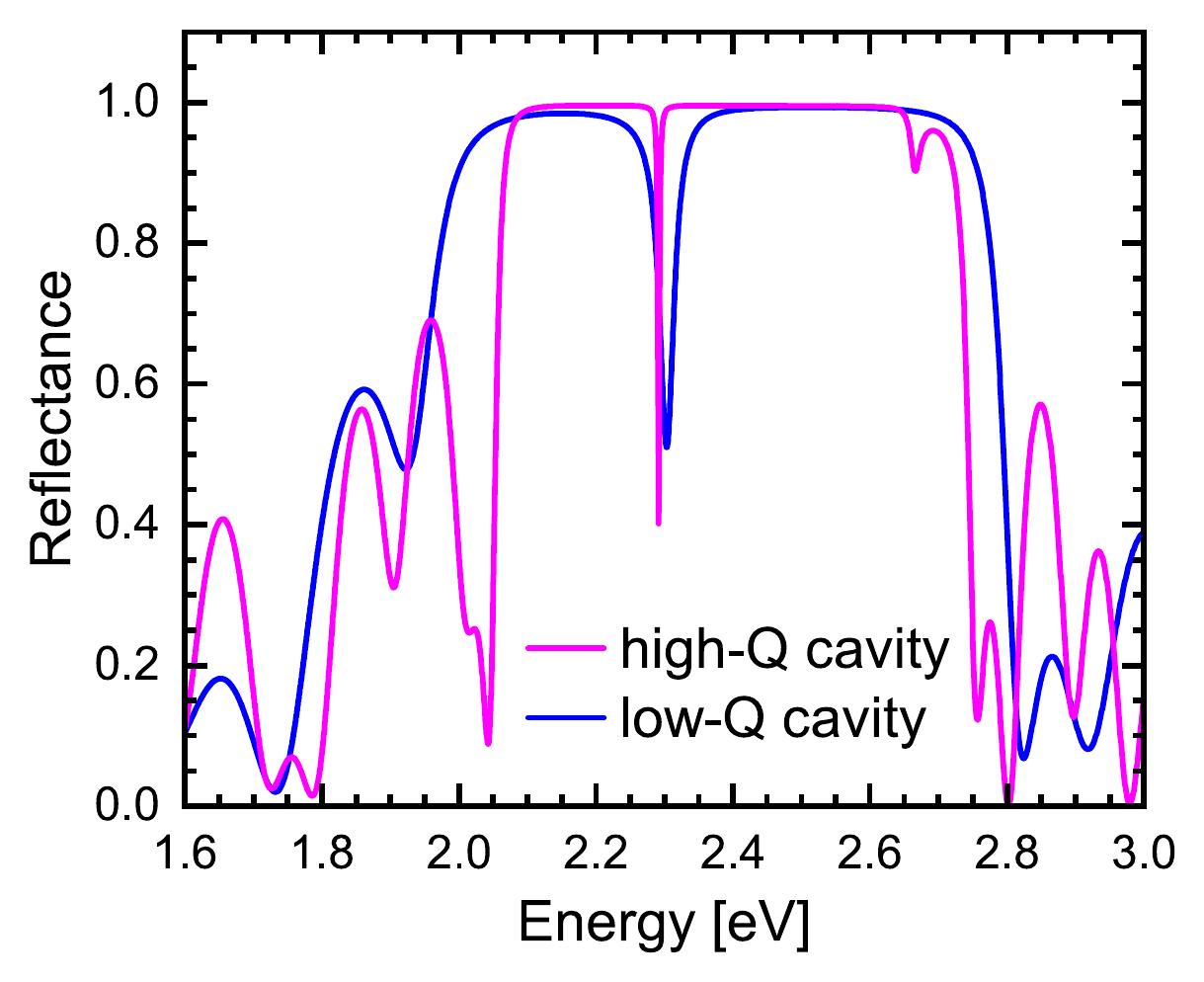}
    \caption{\textbf{Cavity mode linewidths.} Simulated (TMM) reflectance spectra for the low-$Q$ (blue line) and high-$Q$ (magenta line) cavities, with the rubrene vibronic oscillator strengths set to zero. These simulations allow for the extraction of the ``empty" cavity mode linewidths ($\kappa$), which are $\SI{26}{\milli\electronvolt}$ and $\SI{2}{\milli\electronvolt}$ for the low-$Q$ and high-$Q$ cavities, respectively.}
    \label{SI_fig_cavity_mode_linewidth}
\end{figure}

The quality factor ($Q$) of the microcavities was determined using the standard relation:
\begin{equation}
Q = \frac{E_0}{\kappa}
\label{eq:q_factor}
\end{equation}
where $E_0$ is the central resonance energy of the cavity mode at normal incidence and $\kappa$ represents the full-width at half-maximum ($\mathrm{FWHM}$) energy linewidth. 

The corresponding cavity photon lifetime ($\tau_c$), which dictates the characteristic time scale for photon decay within the resonator, is inversely proportional to the energy linewidth via the energy-time uncertainty relation:
\begin{equation}
\tau_c = \frac{\hbar}{\kappa}
\label{eq:cavity_lifetime}
\end{equation}
where $\hbar \approx \SI{6.5821e-16}{\electronvolt\second}$ ($\approx \SI{0.6582}{\electronvolt\femto\second}$) is the reduced Planck constant. Alternatively, the lifetime can be expressed as a function of the quality factor and resonance energy:
\begin{equation}
\tau_c = \frac{\hbar Q}{E_0}
\label{eq:lifetime_via_Q}
\end{equation}

The linewidth of the cavity mode for each system was found through transfer matrix modeling of the respective cavities with the vibronic transition oscillator strengths set to zero. This simulates a cavity mode filled with a material having the background refractive index of rubrene, but without any light-matter coupling. We then read off the $\mathrm{FWHM}$ ($\kappa$) from the simulated reflectance data at normal incidence ($\theta = 0^\circ$), yielding $\kappa = \SI{26}{\milli\electronvolt}$ and $\kappa = \SI{2}{\milli\electronvolt}$ for the low-$Q$ and high-$Q$ cavities, respectively. 

Given a central cavity resonance energy of $E_0 \approx \SI{2.30}{\electronvolt}$, these linewidths yield a calculated quality factor of $Q \approx 90$ and a corresponding photon lifetime of $\tau_c = \frac{\SI{0.6582}{\electronvolt\femto\second}}{\SI{0.026}{\electronvolt}} \approx \SI{25}{\femto\second}$ for the low-$Q$ cavity architecture. For the high-$Q$ cavity architecture, these parameters correspond to $Q \approx 1150$ and a longer photon lifetime of $\tau_c = \frac{\SI{0.6582}{\electronvolt\femto\second}}{\SI{0.002}{\electronvolt}} \approx \SI{330}{\femto\second}$.

\subsection{Open cavity structure and angular reflectance}\label{SI_sec_opencavity}

\begin{figure}[H]
    \centering
    \includegraphics[width=0.9\linewidth]{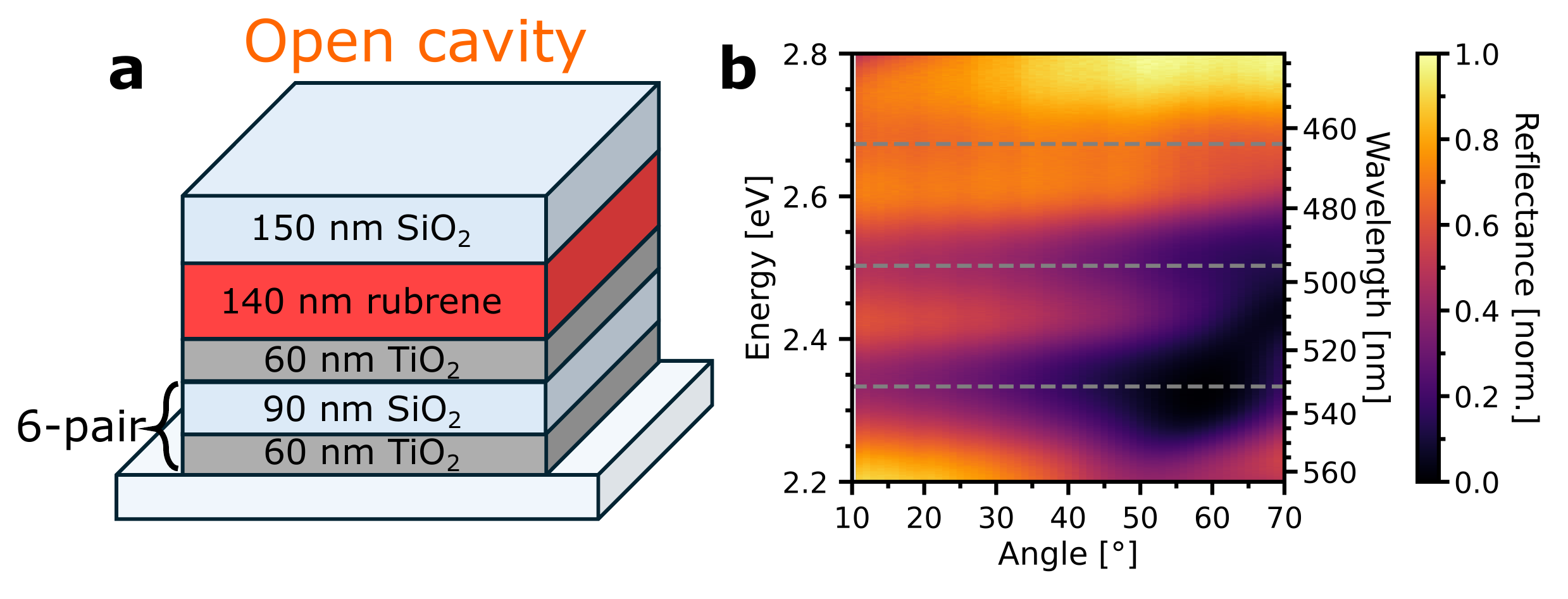}
    \caption{\textbf{Open cavity design and linear characterization.} \textbf{(a)} Structure of the open cavity grown on a quartz-capped glass substrate. The structure consists of a $6$-pair bottom $\mathrm{SiO_2}/\mathrm{TiO_2}$ distributed Bragg reflector (DBR) (the same type of bottom mirror used in the low-$Q$ cavities), followed by a $\SI{140}{\nano\meter}$ thermally evaporated rubrene active layer, and a $\SI{150}{\nano\meter}$ $\mathrm{SiO_2}$ encapsulating layer to prevent degradation. \textbf{(b)} Angle-resolved white light reflectance map of the open cavity, with the energies of rubrene's prominent $\mathrm{S_0 \to S_1}$ vibronic transitions overlaid in dashed grey lines.}
    \label{SI_fig_opencavity}
\end{figure}

In the reflectance maps there is a reduction in reflectance at the energies of rubrene's $\mathrm{S_0 \to S_1}$ vibronic transitions due to the absorption reducing the intensity of reflected light. There also appears to be a reduced reflectance at wider angles, which appears to shift to higher energies with increasing angles. This can be rationalized by the distributed Bragg reflector stopband centre shifting to greater energies with increasing viewing angle, thus the reflectance is reduced in regions outside of the stopband. There is no indication of strong coupling.

\subsection{Strong coupling criterion}
\label{SI_sec_strong_coupling_criterion}

\begin{figure}[H]
    \centering
    \includegraphics[width=0.6\linewidth]{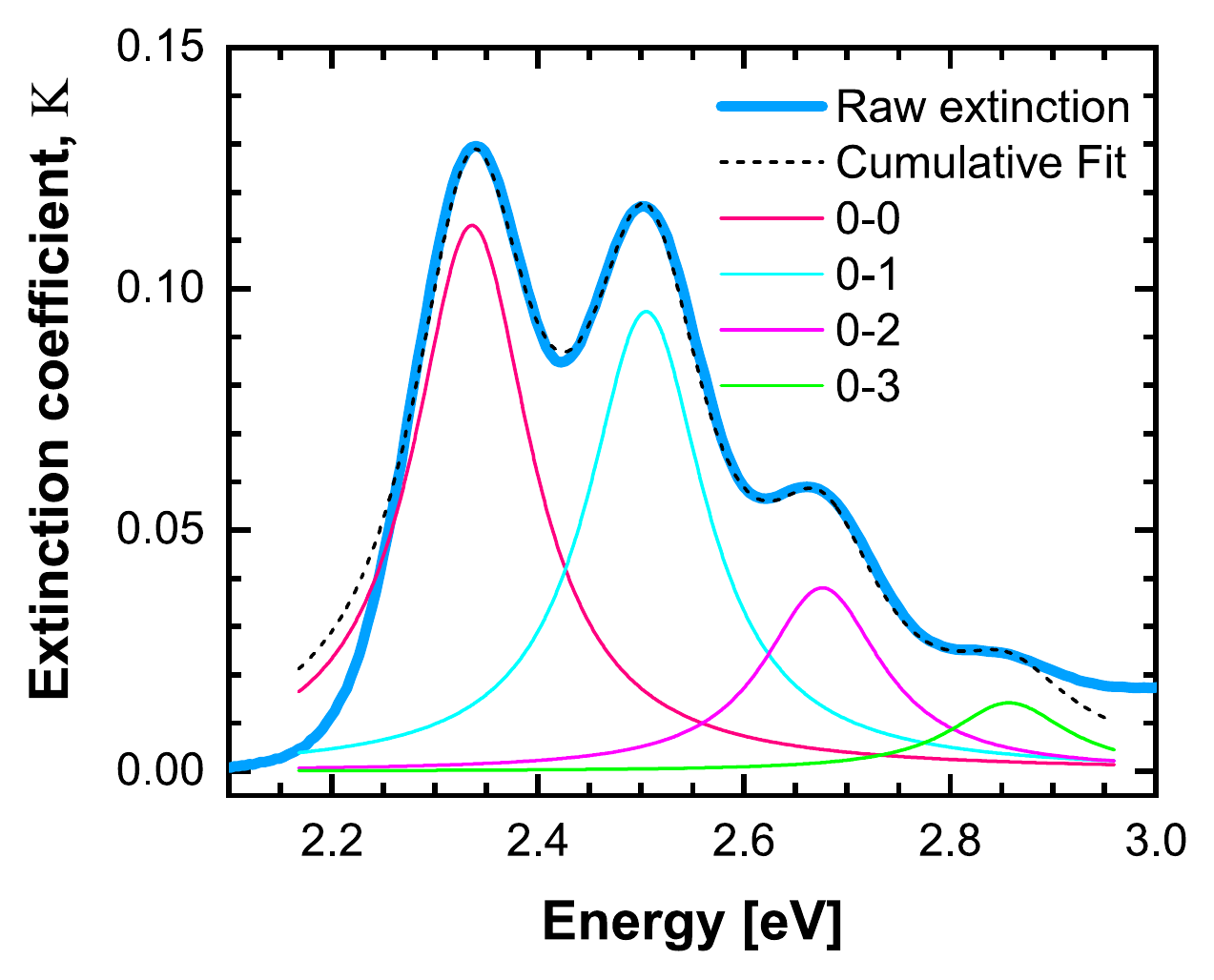}
    \caption{\textbf{Determination of exciton and cavity mode linewidths.} Extinction spectrum of a rubrene film prepared by thermal evaporation, fitted with four Lorentzian peaks (0-0, 0-1, 0-2, 0-3 transitions) representing the vibronic transitions. The cumulative fit (dashed black line) closely matches the raw extinction data (solid blue line), yielding an exciton linewidth ($\Gamma$) of $\SI{130}{\milli\electronvolt}$.}
    \label{SI_fig_strong_coupling_criterion}
\end{figure}

In order to calculate the strong coupling threshold for both cavities, we require the linewidth ($\Gamma$) of each vibronic transition and the linewidth ($K$) of the cavity mode. The linewidths of the rubrene vibronic transitions were determined by fitting the extinction spectrum with a series of three Lorentzians centred at \SI{2.33}{\electronvolt}, \SI{2.50}{\electronvolt}, and \SI{2.67}{\electronvolt}, respectively. During this fitting process, the linewidth for all transitions was constrained to a single value, yielding a full width at half maximum ($\mathrm{FWHM}$) of $\SI{130}{\milli\electronvolt}$. Extraction of the cavity mode linewidths is shown in \ref{SI_sec_cavity_mode_linewidth}, yielding \SI{26}{\milli\electronvolt} and \SI{2}{\milli\electronvolt} for the low-Q and high-Q microcavities, respectively. The strong-coupling threshold for each cavity type can then be calculated using the following formula:

\begin{equation}
\hbar\Omega = \frac{\Gamma + K}{2}.
\end{equation}
This provides us with a strong coupling threshold of \SI{78}{\milli\electronvolt} and \SI{66}{\milli\electronvolt} for the low-Q and high-Q microcavities, respectively.

\subsection{Rate equations and constants}\label{SI_sec_kinetic_model} 

Transient absorption kinetics were modeled using the kinetic scheme outlined in Figure \ref{fig_SF_TA_kinetics}d of the main text, which was adapted from a previously published model.\cite{bossanyiOptimizedRubrenebasedNanoparticle2022} Here the time-dependent populations of the singlet $\mathrm{[S_1]}$, strongly exchanged-exchange coupled singlet triplet-pair $\mathrm{[^1(TT)]}$, the weakly exchanged-coupled singlet triplet-pair $\mathrm{[^1(T...T)]}$, and the mixed spin triplet-pair $\mathrm{[(T...T)}^l]$ were simulated using the following rate equations:

\begin{align}
\frac{d[\text{S}_1]}{dt} &= -(k_r + k_{sf}) [\text{S}_1] - k_{ssa} [\text{S}_1]^2 + k_{tf} [\text{}^1\text{(TT)}] \\
\frac{d[^1\text{(TT)}]}{dt} &= k_{sf} [\text{S}_1] - (k_{hop} + k_{tf}) [^1\text{(TT)}] + k_{-hop} [^1(\text{T}\dots\text{T})] \\
\frac{d[^1(\text{T}\dots\text{T})]}{dt} &= k_{hop} [^1\text{(TT)}] - (k_{-hop} + k_{spin} + k_t) [^1(\text{T}\dots\text{T})] \\
\frac{d[(\text{T}\dots\text{T})]}{dt} &= k_{spin} [^1(\text{T}\dots\text{T})] - k_{t} [(\text{T}\dots\text{T})]
\end{align}

The values of the rate constants for each sample type are listed in Table \ref{SI_tab_rate_constants}, and their definitions can be found in the original publication of the model.\cite{bossanyiOptimizedRubrenebasedNanoparticle2022} Their values were optimized by manually fitting the kinetics to the measured data using home-built python code and minimizing the residuals (Figure \ref{SI_fig_kinetics_residuals} of the SI).

We find a best fit to the kinetics when modeling both the closed cavities (coupled) and, importantly, the open cavity control (uncoupled) with an $\sim\SI{20}{\percent}$ enhancement of the first ($k_{\mathrm{sf}}$) and second step ($k_{\mathrm{hop}}$) of SF when compared to the bare film. Although it is tempting to assign the enhanced SF rates to strong-coupling, as predicted by some theoretical models when exclusively exciting the LPB, \cite{martinez-martinezPolaritonAssistedSingletFission2018, climentNotDarkStrong2022} the open-cavity control sample also exhibits the enhanced SF rates. This implies that the enhancement instead derives from the surface contact between rubrene and the adjacent $\text{TiO}_2$ layer of the bottom DBR. As $\text{TiO}_2$ can be used as an electron transport layer,\cite{qiuElectronBeamEvaporated2015} and has been shown to modulate SF with other acenes when covalently linked to $\text{TiO}_2$ nanoparticles,\cite{paceDynamicsSingletFission2018} it is plausible that the exciton dynamics are altered slightly at the rubrene--$\text{TiO}_2$ interface, explaining the discrepancy between the two uncoupled controls.

\begin{table}[h]
    \centering
    \caption{Optimized values and sources of the rate constants used for modeling the dynamics of the rubrene film and cavities (open cavity, low-$Q$ cavity, and high-$Q$ cavity).}
    \label{SI_tab_rate_constants}
    \renewcommand{\arraystretch}{1.2} 
    
    \begin{tabular}{l l l l}
        \toprule
        \textbf{Rate} & \textbf{Film} & \textbf{Cavity} & \textbf{Source} \\
        \midrule
        $k_{sf}$ & $\SI{110}{ns^{-1}}$ & $\SI{130}{ns^{-1}}$ & TA data, then optimised \\
        $k_{tf}$ & $\SI{180}{ns^{-1}}$ & $\SI{170}{ns^{-1}}$ & TA data, then optimised \\
        $k_{hop}$ & $\SI{8}{ns^{-1}}$ & $\SI{10}{ns^{-1}}$ & Ref.\cite{bossanyiOptimizedRubrenebasedNanoparticle2022}, then optimised \\
        $k_{-hop}$ & $\SI{0.7}{ns^{-1}}$ & $\SI{0.7}{ns^{-1}}$ & Ref.\cite{bossanyiOptimizedRubrenebasedNanoparticle2022}, then optimised \\
        $k_{spin}$ & $\SI{0.25}{ns^{-1}}$ & $\SI{0.25}{ns^{-1}}$ & MFE data, Ref.\cite{wolfQuantumBeatsMultiexciton2018}, fixed \\
        $k_{ssa}$ & $\SI{1.4e-17}{cm^{3}ns^{-1}}$ & $\SI{2e-17}{cm^{3}ns^{-1}}$ & Ref.\cite{bossanyiOptimizedRubrenebasedNanoparticle2022}, then optimised \\
        $k_{r}$ & $\SI{0.0625}{ns^{-1}}$ & $\SI{0.0625}{ns^{-1}}$ & Ref.\cite{bossanyiOptimizedRubrenebasedNanoparticle2022}, fixed \\
        $k_{t}$ & $\SI{1e-5}{ns^{-1}}$ & $\SI{1e-5}{ns^{-1}}$ & Ref.\cite{ryasnyanskiyTripletExcitonDynamics2011a}, fixed \\
        \bottomrule
    \end{tabular}
\end{table}

The model also includes a singlet-singlet exciton annihilation term which is population dependent. This requires the initial photo-excited singlet population to accurately fit the data, which can be calculated using the following equation:

\begin{equation}
N_0 \approx \frac{P}{f A} \frac{\lambda}{hc} \frac{1}{d} (1 - 10^{-OD})
\label{eq_excitation_density}
\end{equation}

The initial singlet population density, $N_0$, is determined by several experimental and material parameters. Here, $P$ is the measured pump power, $f$ is the repetition rate of the excitation laser (\SI{10}{kHz}), and $A$ is the beam spot area. The excitation wavelength ($\lambda = \SI{530}{nm}$) is used to calculate the energy of the incident pump photons via Planck's constant, $h$, and the speed of light, $c$. Knowledge of the sample's absorption properties is required, which are determined by the sample thickness, $d$, and the pump absorbance, $\text{OD}$. For the film, an absorbance of \SI{0.13}{\text{OD}} was used.

However, conventional absorbance measurements of the cavities are complicated by the presence of their mirrors. To address this problem, we opted to estimate the absorbance of the cavities using their initial singlet ($\SI{1180}{}-\SI{1220}{nm}$) $\Delta A$ signal measured at low fluence. As $\Delta A$ is directly proportional to the pump absorbance, this method allows us to directly measure the absorbance of the cavities by comparing to the initial $\Delta A$ signal of the film. This is acceptable because we have already established in the main text that the near-infrared (NIR) singlet transition of the cavities is mostly protected from pump-induced artefacts. Here we find pump absorbances of $\text{OD} = 0.17$, $\text{OD} = 0.24$, and $\text{OD} = 0.029$ for the open cavity, low-$Q$ cavity, and high-$Q$
cavity, respectively. From these absorbance values, we are able to calculate the initial singlet population present for each of the samples, which we then implement in our model at the different pump pulse energies specified in the relevant figures.

It is worth noting here that the transient absorption measurements presented in this publication were deliberately taken at low excitation densities to avoid significant singlet-singlet exciton annihilation. Therefore, any error in the calculated excitation densities will not have a significant impact on the modeling of second order dynamics.

\subsection{Time-dependent cavity spectra}\label{SI_sec_cavity_spectra}

\begin{figure}[H]
    \centering
    \includegraphics[width=0.9\linewidth]{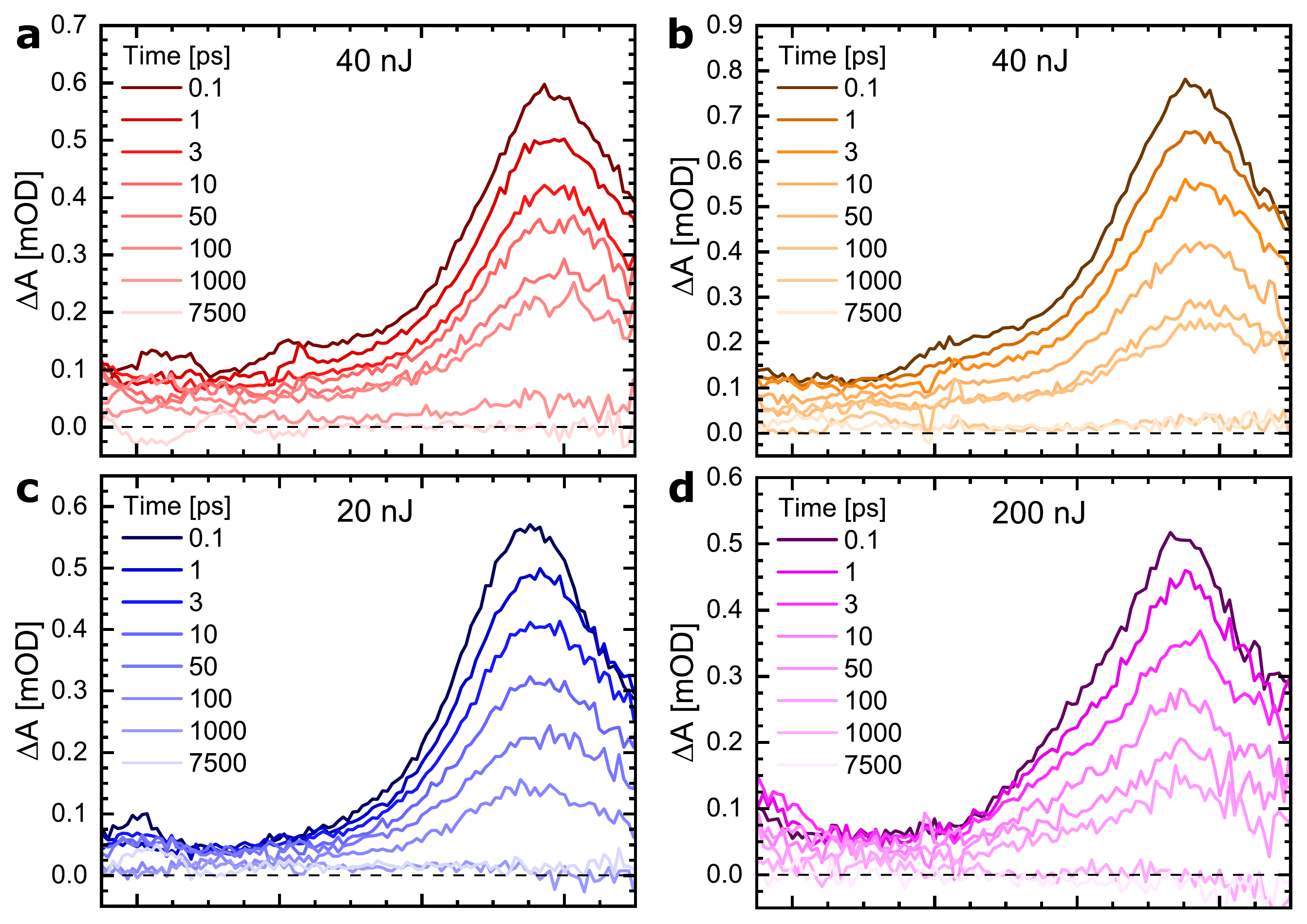}
    \caption{\textbf{Transient absorption spectra of rubrene microcavities and film.} NIR transient absorption ($\Delta A$) spectra at various delay times for four samples following $\SI{530}{\nano\meter}$ excitation: (\textbf{a}) bare film, (\textbf{b}) open-cavity, (\textbf{c}) low-Q cavity, and (\textbf{d}) high-Q cavity. The incident pump-fluence was set at $\SI{8}{\micro\joule\per\centi\meter\squared}$, $\SI{4}{\micro\joule\per\centi\meter\squared}$, $\SI{8}{\micro\joule\per\centi\meter\squared}$ and $\SI{72}{\micro\joule\per\centi\meter\squared}$, respectively, to maintain a similar initial peak $\Delta A$ signal ($\SI{0.5}{}$--$\SI{0.8}{mOD}$). The pumps pulse energy is given at the top of each panel.}
    \label{SI_fig_cavity_spectra}
\end{figure}

Examining the time-dependent spectra, we see a rapid decay of the $\SI{1200}{\nano\meter}$ singlet peak relative to the $\SI{850}{}\text{--}\SI{1050}{\nano\meter}$ triplet band, indicating that triplets are being generated by SF at similar rates for each of the samples.

\subsection{Transient absorption TMM simulations}\label{SI_TA_TMM}

The presence of artifacts in the NIR is plausible given that the reflectance of the $\mathrm{DBR}$s in the $\mathrm{NIR}$ region is non-zero. We are thus prompted to understand these spectral changes in more detail, accounting for potential pump-induced spectral artifacts \cite{renkenUntargetedEffectsOrganic2021} by simulating the $\mathrm{TA}$ spectra using a $\mathrm{TMM}$ approach.

We use the $\mathrm{TMM}$ to simulate the constituent parts of the $\mathrm{TA}$ measurements. In $\mathrm{TA}$ a ``pump-on" excited-state transmittance ($T_{\text{on}}$) is compared with a ``pump-off" ground-state transmittance ($T_{\text{off}}$) to calculate the change in absorbance:

\begin{equation}
\Delta A = - \log_{10} \left( \frac{T_{\text{on}}}{T_{\text{off}}} \right)
\label{eq:dA_deltaA}
\end{equation}

\begin{figure}[H]
    \centering
    \includegraphics[width=1\linewidth]{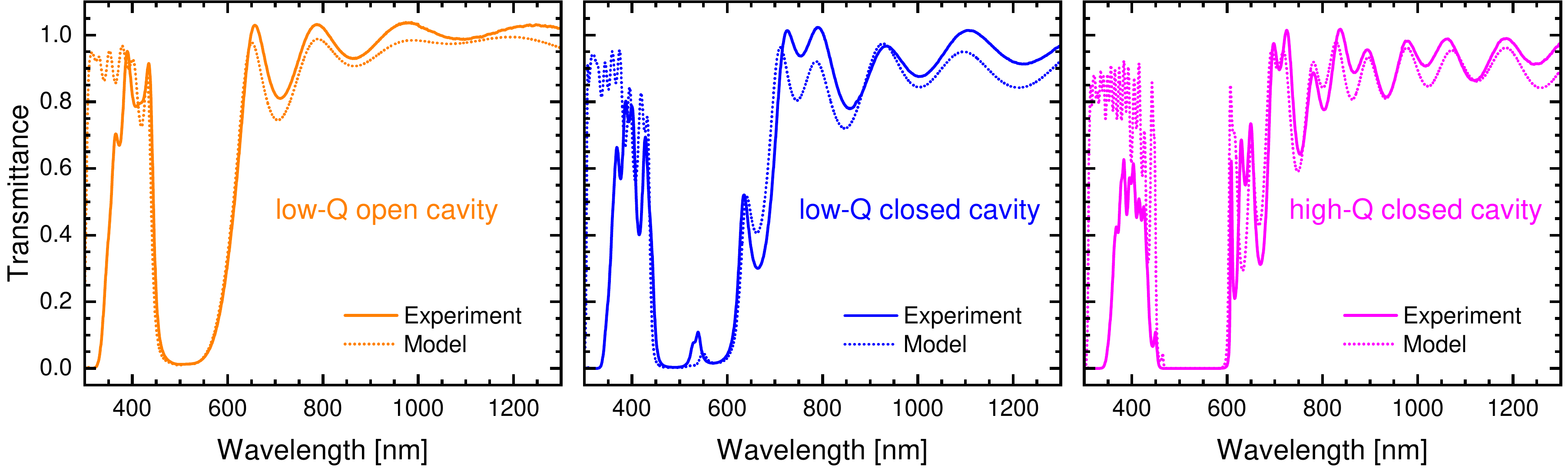}
       \caption{\textbf{Simulated and measured cavity transmittance at normal incidence.} The measured transmittance spectra (solid lines) are accurately reproduced by the transfer-matrix method (dotted lines), thereby validating the optical parameters used for the subsequent $\Delta A$ simulations. The shape of the Fabry-Perot interference fringes are successfully captured, though there is some residual error presumably due to the model assuming perfectly equal $\text{SiO}_2$ or $\text{TiO}_2$ thicknesses throughout the DBRs.}
    \label{SI_fig_cav_transmission}
\end{figure}

A proxy $\mathrm{TA}$ measurement is possible by using the $\mathrm{TMM}$ to simulate these ground-state and excited-state transmittance's while taking into account the optical response of the $\mathrm{DBR}$s. The pump-off optical response of the microcavities is achieved in the cavity design stage: the film's extinction spectrum was fitted with Lorentzian oscillators and fed into the $\mathrm{TMM}$ alongside the optical properties of the DBR layers to simulate the angular reflectance of the cavities as shown in the main text. For the purpose of modeling the TA data, we have extended the pump-off transmittance simulation to the NIR region, as shown in Figure \ref{SI_fig_cav_transmission}.

\begin{figure}[H]
    \centering
    \includegraphics[width=0.7\linewidth]{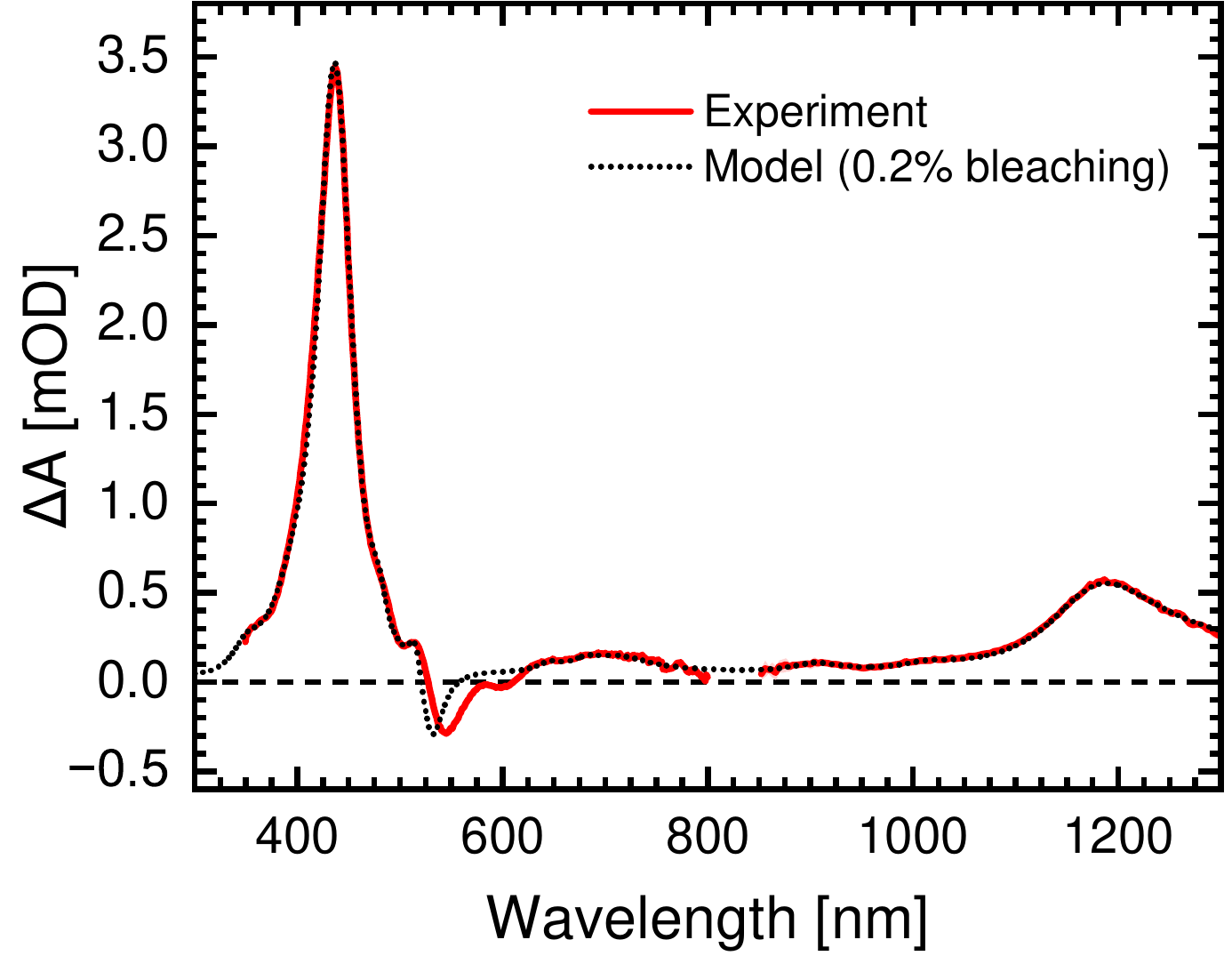}
       \caption{\textbf{Transient absorption simulation of a thermally evaporated rubrene film at normal incidence.} The experimental spectrum was taken at a $\SI{0.1}{}$--$\SI{0.3}{ps}$ pump-probe delay upon $\SI{530}{\nano\meter}$ excitation at $\SI{8}{\micro\joule\per\centi\meter\squared}$ fluence. This data was modelled by simulating the ground-state transmittance and manually optimising the photo-induced absorption through the transfer matrix method before calculating the $\Delta A$ signal. This model assumes $\SI{0.2}{\percent}$ of the ground-state population is bleached upon photo-excitation. We see an almost identical match between the model and simulation -- only the bleaching region ($\sim\SI{530}{\nano\meter}$) is not perfectly described by the model due to overlap with the photo-induced absorption.}
    \label{SI_fig_film_sim_TA}
\end{figure}

For the pump-on transmittance, we first model the bare film before incorporating the mirrors. This involves adding new Lorentzian oscillators to the $\mathrm{TMM}$ corresponding to the rubrene $\mathrm{ESA}$ peaks and reducing the ground-state absorption oscillator strength to capture the effect of bleaching. The $\mathrm{ESA}$ spectrum of the film is then calculated using Equation \ref{eq:dA_deltaA}, where the final $\Delta A$ spectrum is manually optimised by adjusting the pump-on transmittance parameters. The resulting optimised proxy $\mathrm{TA}$ measurement for the film, shown in Figure \ref{SI_fig_film_sim_TA}, exhibits an almost perfect match to the experimental data.

\textbf{}

\begin{figure}[H]
    \centering
    \includegraphics[width=0.7\linewidth]{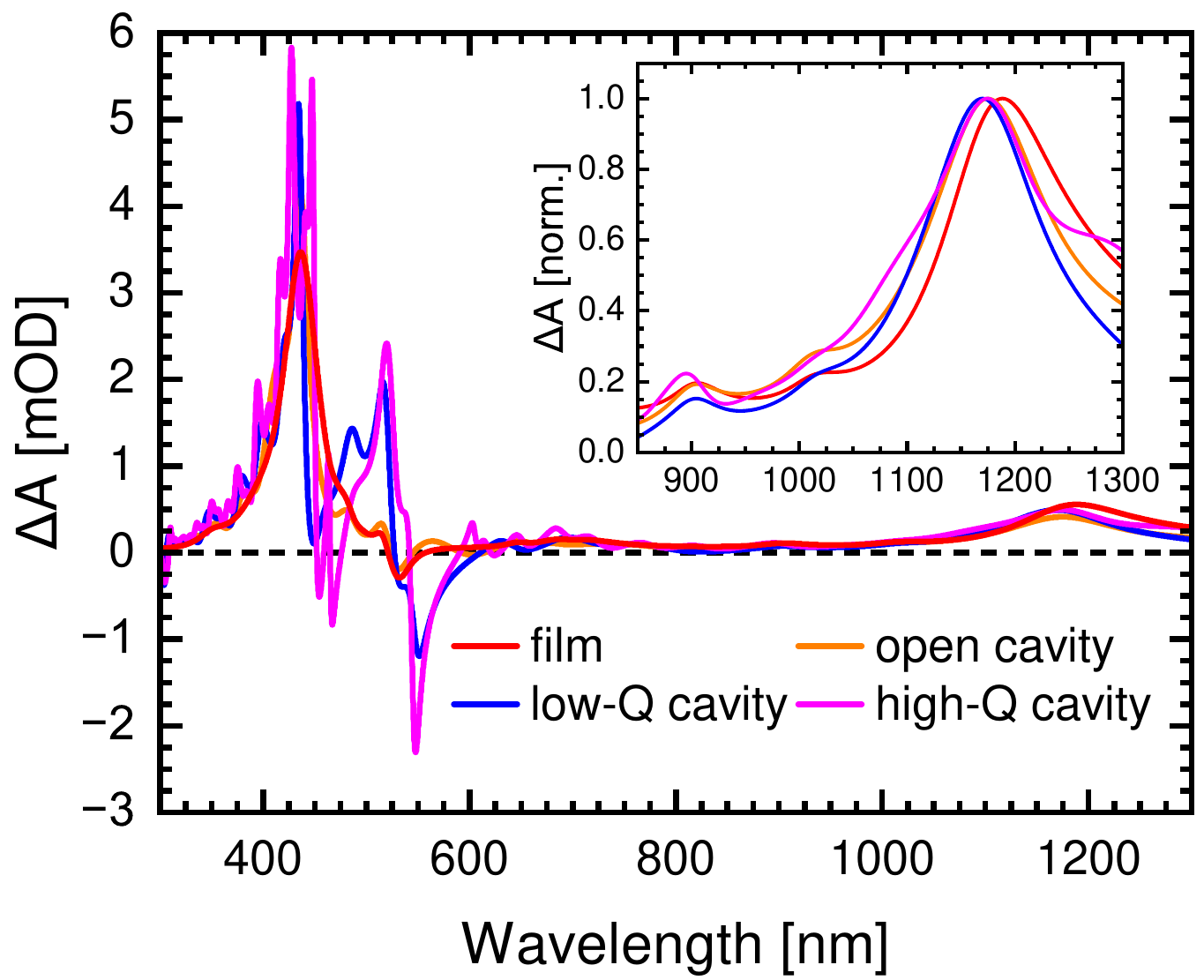}
       \caption{\textbf{Transient absorption simulation of the rubrene microcavities and film at normal incidence.} The transfer-matrix method simulation is based on the measured $\SI{0.1}{}$--$\SI{0.3}{ps}$ transient absorption signal of the film and assumes $\SI{0.2}{\percent}$ of the ground-state population is bleached upon photo-excitation. The inset shows the $\SI{850}{}$--$\SI{1300}{nm}$ spectra normalized to the maximum signal.}
    \label{SI_fig_cavity_sim_TA}
\end{figure}

We now incorporate the DBR layers into the simulation to calculate the pump-on transmittance of the cavities before calculating the $\Delta A$ signal (using SI Equation \ref{eq:dA_deltaA}) for the various structures, as presented in Figure \ref{SI_fig_cavity_sim_TA}. We simulated the $\SI{0.1}{}$--$\SI{0.3}{\pico\second}$ spectra as this time window has the highest signal-to-noise ratio and allows direct comparison with the prompt spectra presented in the main text.

As Figure \ref{SI_fig_cavity_sim_TA} shows, the simulation recovers the visible response of the microcavities despite our experiment only probing the $\mathrm{NIR}$ region. Upon comparison of the two spectral ranges, it is clear that new, dominating features emerge in the visible region with the closed cavities, but the NIR region remains mostly similar to the bare film.

However, there are some subtle spectral deviations in the NIR likely due to additional pump-induced effects contributing to the measured $\mathrm{TA}$ signal, including thermal expansion, changes in the background refractive index of the $\mathrm{DBR}$s/active layer, or varying magnitudes of ground-state bleaching. We systematically investigate these pump-induced effects by simulating the $\mathrm{TA}$ spectra in the visible and, critically, the $\mathrm{NIR}$ regions with our model for the open cavity, low-Q cavity, and high-Q cavity. Given that there is only a minor deviation of the cavities' $\mathrm{S_{1} \to S_{N}}$ ($\SI{1200}{\nano\meter}$) peak relative to the film in the experiment, we have chosen to simulate pump-induced effects using a parameter range that causes only a small perturbation of this peak.

\begin{figure}[H]
    \centering
    \includegraphics[width=1\linewidth]{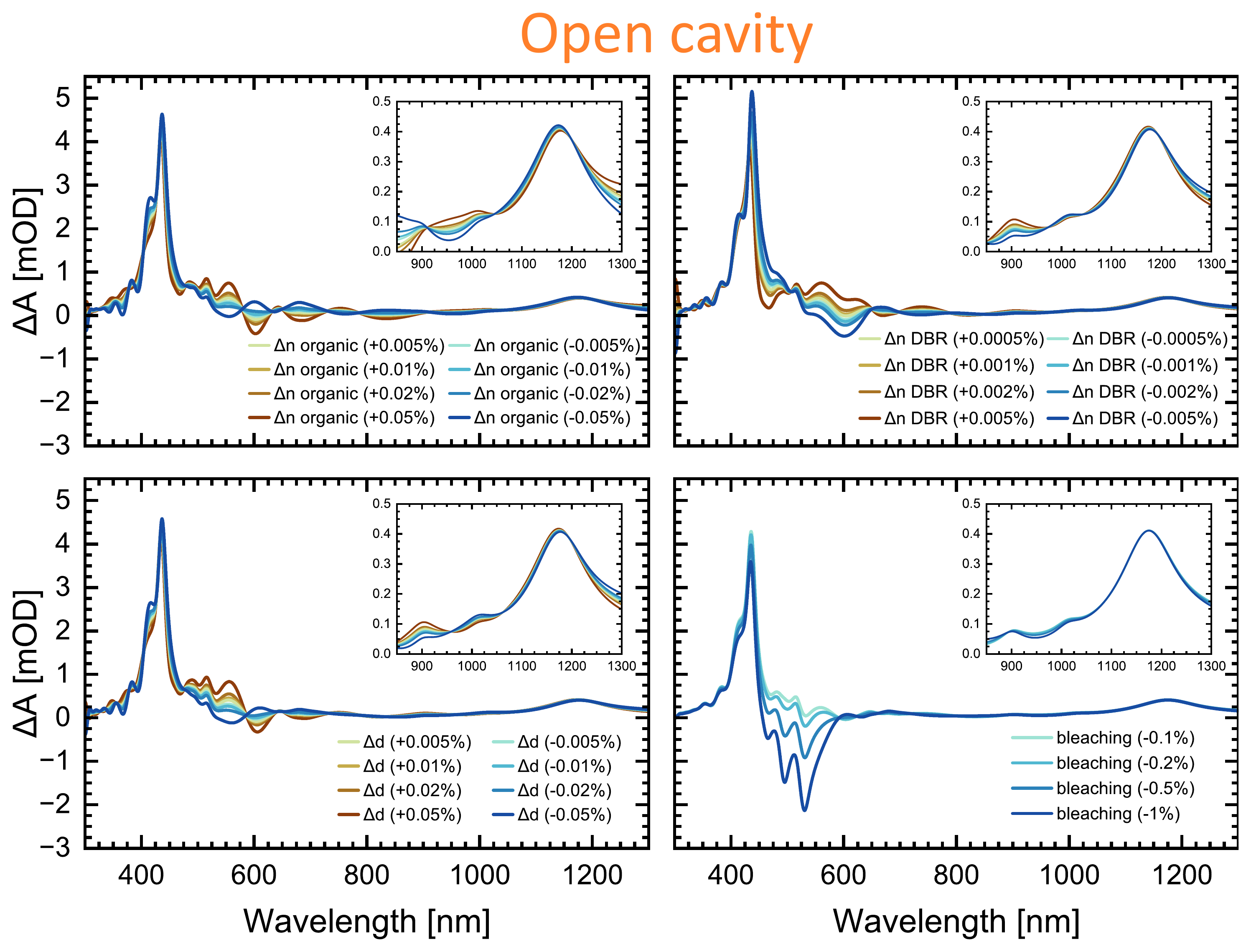}
        \caption{\textbf{Open cavity transient absorption simulations at normal incidence.} The transfer matrix simulations, taken at $\SI{0.1}{}$--$\SI{0.3}{ps}$ delay, show the change in the $\Delta A$ signal when individually varying the following photo-induced effects, presented as a percentage change from the ground-state values: (a) Background refractive index of the rubrene layer, (b) background refractive index of the $\mathrm{DBR}$ mirror layers ($\mathrm{TiO_2}$ and $\mathrm{SiO_2}$), scaled by the same percentage, (c) change in thickness of the rubrene layer (and reduction in oscillator strengths to keep the number of absorbers constant), and (d) ground-state bleach.}
    \label{SI_fig_sim_TA_artefacts_open_cavity}
\end{figure}

\begin{figure}[H]
    \centering
    \includegraphics[width=1\linewidth]{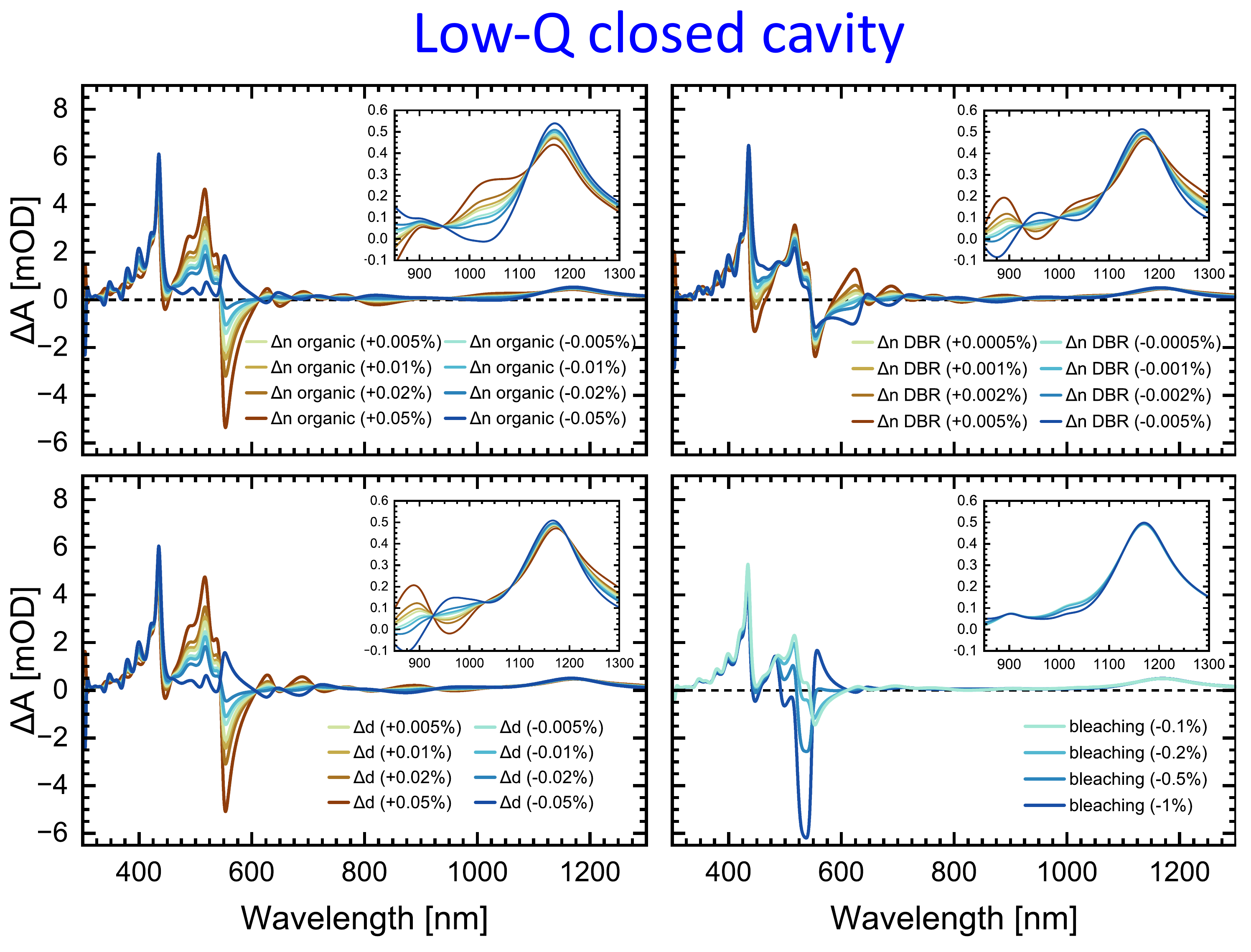}
    \caption{\textbf{Low-Q cavity transient absorption simulations at normal incidence.}}
    \label{SI_fig_sim_TA_artefacts_lowQ_cavity}
\end{figure}

\begin{figure}[H]
    \centering
    \includegraphics[width=1\linewidth]{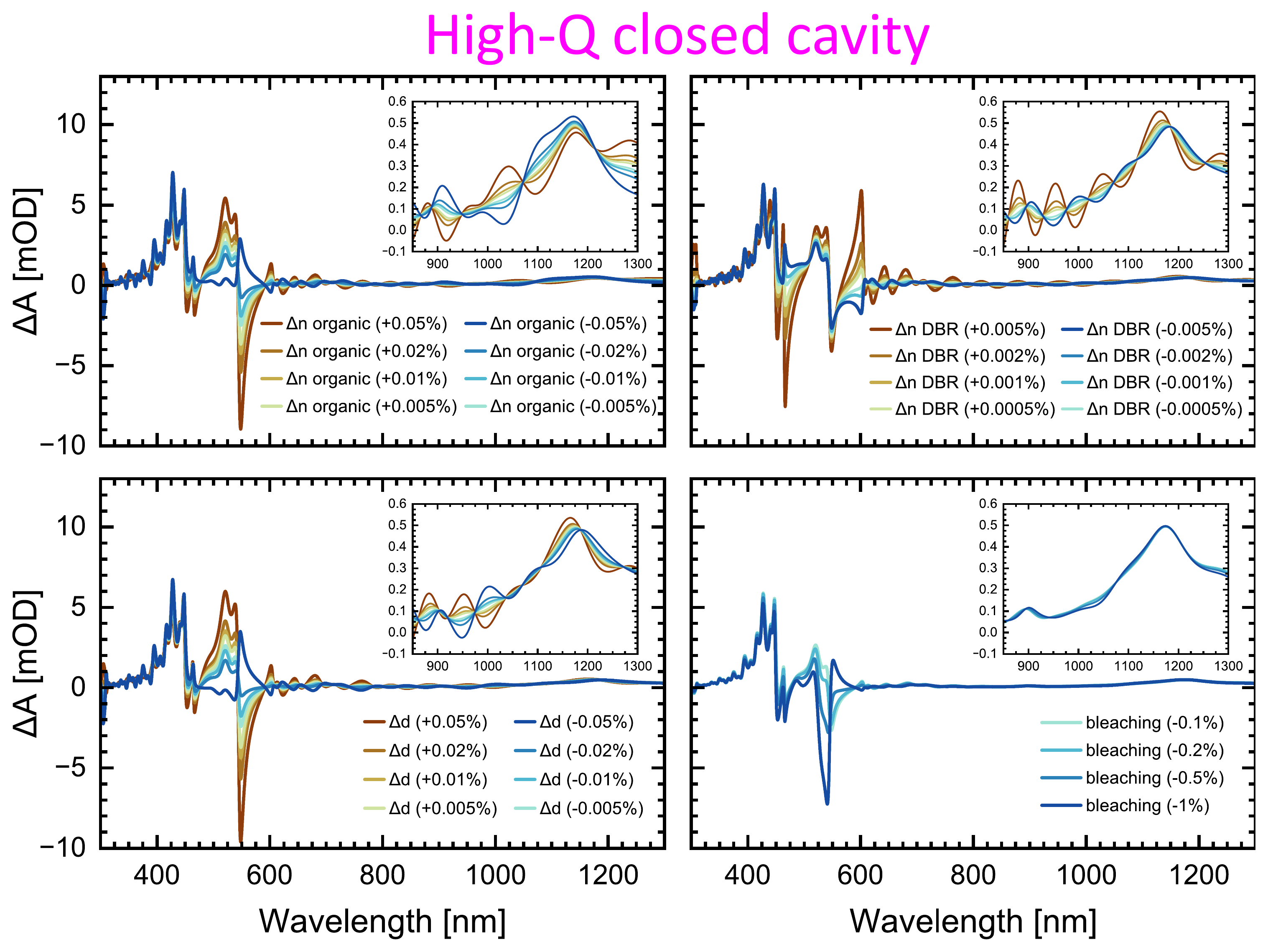}
    \caption{\textbf{High-Q cavity transient absorption simulations at normal incidence.}}
    \label{SI_fig_sim_TA_artefacts_highQ_cavity}
\end{figure}

Upon examining the simulated pump-induced effects of the closed microcavities, it is clear that major artifacts are again produced in the visible region due to the high reflectivity of the $\mathrm{DBR}$s and the cavity mode supported there, but the NIR region is mostly protected from these pump-induced effects. Assessing the NIR artifacts more closely we find the high-$\mathrm{Q}$ cavity is more prone to cavity-induced artifacts due to the high reflectivity mirrors spectrally modulating the Fabry-Perot interference fringes more frequently in the NIR region, as shown in SI Figure \ref{SI_fig_cav_transmission}. In contrast, the open cavity exhibits only very subtle spectral deviations across the entire wavelength range, suggesting that the optical confinement imposed by both mirrors is the primary condition for pump-induced artifacts.

\begin{figure}[H]
    \centering
    \includegraphics[width=1\linewidth]{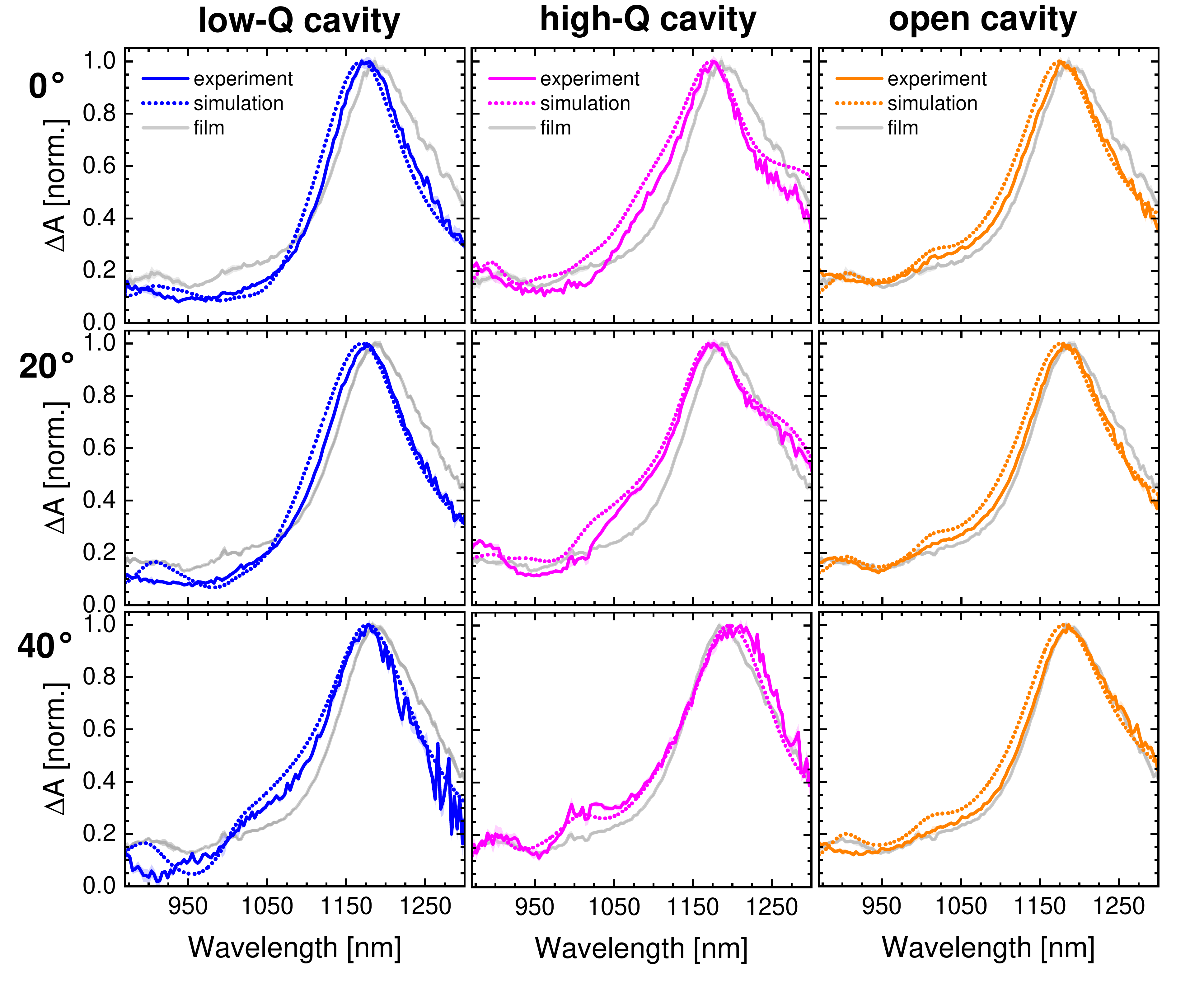}
    \caption{\textbf{Optimized angle-dependent transient absorption simulations.} The $\Delta A$ spectra of the different cavities are simulated (dotted line) at incidence angles of $\SI{0}{\degree}$, $\SI{20}{\degree}$ and $\SI{40}{\degree}$ using the pump-induced changes (relative to the ground-state) listed in Table \ref{tab_optimised_angular_params}. We also overlay the respective experimental spectra of the cavities (solid line) and film (grey line) at the same angles of incidence. All spectra were either simulated or measured at a pump-probe delay of $\SI{0.1}{}$--$\SI{0.3}{ps}$.}
    \label{fig_cavity_TA_spectra_angular_modelling}
\end{figure}

Having systematically investigated the pump-induced artifacts independently, we now use these simulations to guide our data modeling. We apply realistic and optimized combinations of the pump-induced parameters to simulate the $\mathrm{TA}$ spectra at various angles of incidence ($\SI{0}{\degree}$, $\SI{20}{\degree}$, $\SI{40}{\degree}$) in the $\mathrm{NIR}$ region. We then directly compare the simulation with the measured $\mathrm{TA}$ data taken at the same angles, as shown in Figure \ref{fig_cavity_TA_spectra_angular_modelling}. As both the stop-band and Fabry-Perot interference fringes of the $\mathrm{DBR}$s depend on the incident angle, examining the angle-dependent $\mathrm{TA}$ spectra will thoroughly test our model, allowing us to verify whether it accurately describes the measured data.

Table \ref{tab_optimised_angular_params} details the optimized fitting parameters used to model the $\mathrm{TA}$ data. Our physical assumption is that thermal expansion and the associated thermo-optic effect are proportional to the absorbed pump energy (\emph{i.e.} larger initial $\Delta A$), leading to a minor increase in $\Delta d \mathrm{_{org}}$ and a concomitant reduction in $\Delta n \mathrm{_{org}}$.\cite{shengActiveControlElectromagnetic2015} The extent of ground-state bleach was also assumed to be proportional to the pump absorption, and hence the initial $\Delta A$ magnitude. $\Delta n \mathrm{_{DBR}}$, on the other hand, was treated as a free fitting parameter due to the complexity of thermal effects within the multilayer stack. Each of these pump-induced parameters were carefully optimized by hand based on the above assumptions, yielding values that are extremely similar to those previously reported.\cite{renkenUntargetedEffectsOrganic2021}

\begin{table}[h!]
    \centering
    \caption{\textbf{Optimized pump-induced changes used for transfer matrix modeling of the angular transient absorption spectra.} The parameters (listed as a percentage change from ground-state values) are determined by manually fitting the measured data.}
    \label{tab_optimised_angular_params}
    \renewcommand{\arraystretch}{1.2}
    \begin{tabular}{l c c c c}
        \toprule
        Sample & $\Delta n \mathrm{_{org}}$ ($\%$) & $\Delta d \mathrm{_{org}}$ ($\%$) & $\Delta n \mathrm{_{DBR}}$ ($\%$) & Bleaching ($\%$) \\
        \midrule
        Film & $-0.018$ & $0.010$ & $\hphantom{-}0$ & $-0.20$ \\
        Open cavity & $-0.020$ & $0.012$ & $\hphantom{-}0$ & $-0.24$ \\
        Low-$\mathrm{Q}$ cavity & $-0.018$ & $0.010$ & $-0.0015$ & $-0.20$ \\
        High-$\mathrm{Q}$ cavity & $-0.015$ & $0.008$ & $-0.0010$ & $-0.17$ \\
        \bottomrule
    \end{tabular}
\end{table}

Using these optimized parameters, we find that the overwhelming majority of the spectral features are accurately recovered by the model, even at the wider angles of incidence. The features the model cannot reproduce, such as subtle shifts in the $\SI{1200}{\nano\meter}$ peak and minor deviations in the $\SI{850}{}\text{--}\SI{1050}{\nano\meter}$ band, must be associated with the Fabry-Perot interference fringes. This residual error is plausible given the number of assumptions and steps required to model the final TA spectra.

Considering we can reproduce the majority of the spectral changes with the TMM simulation alone heavily suggests that they do not stem from changes to the photophysics of rubrene, but rather pump-induced artifacts.

\subsection{Constituent low-fluence singlet dynamics and residuals}\label{SI_constituent_singlet_dynamics}

\begin{figure}[H]
    \centering
    \includegraphics[width=1\linewidth]{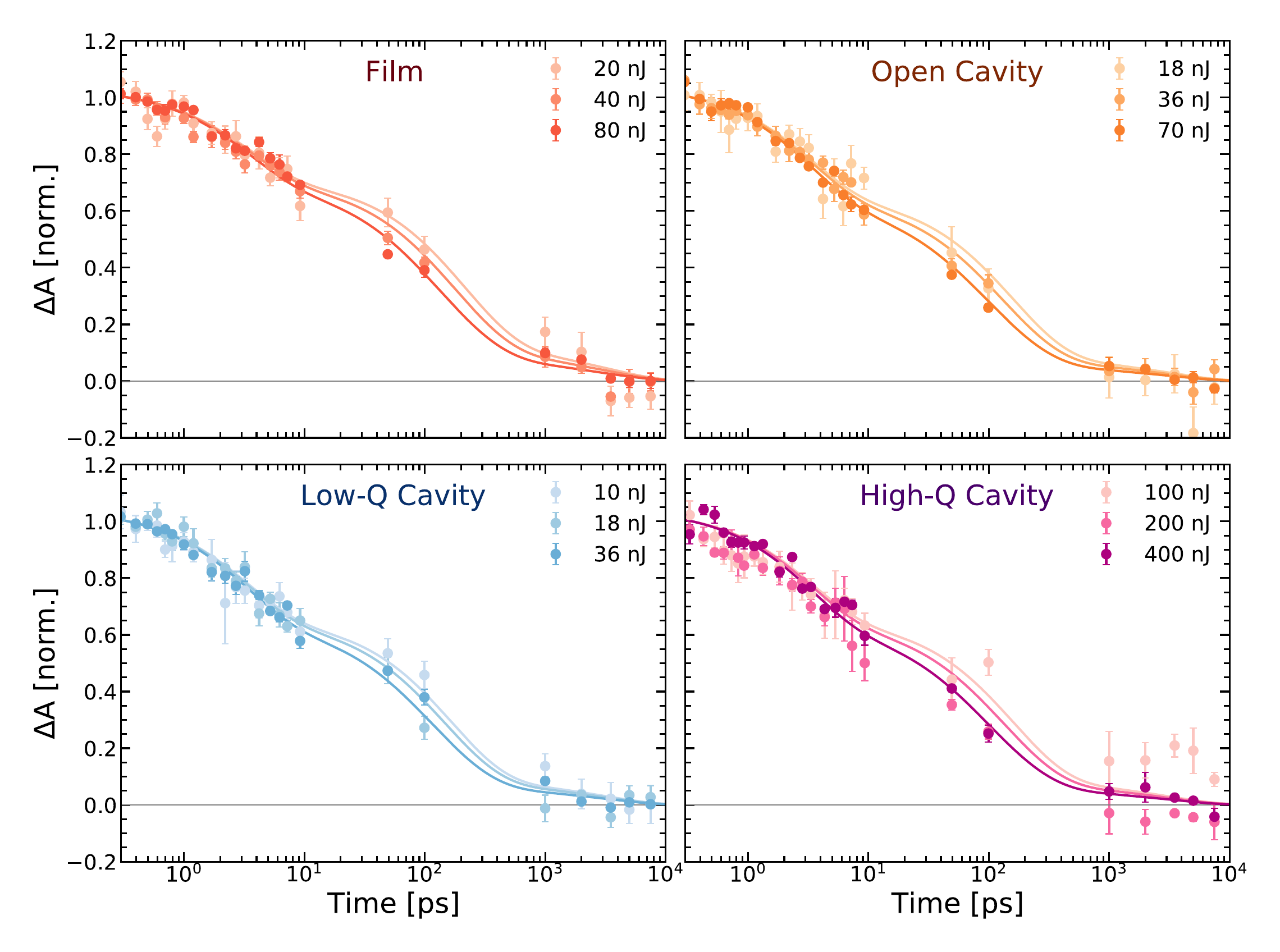}
       \caption{\textbf{Transient absorption singlet dynamics of the rubrene cavities and film.} The kinetics are normalized to the average signal between $\SI{0.2}{}$--$\SI{0.5}{ps}$. Each experimental data point is the average of multiple repeat scans, where the error bar is the standard error of the mean singlet ($\SI{1180}{}$--$\SI{1220}{nm}$) signal of the repeats. The pump pulse energies used for each sample are listed in their respective legend. A beam spot size of $\SI{0.48}{\milli\meter\squared}$ (1/$e^2$) was used for the film, open-cavity and low-$Q$ cavity, whereas a beam spot size of $\SI{0.35}{\milli\meter\squared}$  was used for the high-$Q$ cavity. The kinetic model is overlaid (solid line) for each pulse energy.}
    \label{SI_fig_kinetics}
\end{figure}

The kinetic model used to describe the film, open cavity, low-Q cavity and high-Q cavity data includes a population-dependent singlet-singlet exciton annihilation term, resulting in the dynamics depending on the pump-pulse energy and excitation density. To increase the signal-to-noise ratio, the average of the three low-fluence kinetics were taken, as presented in Figure \ref{SI_fig_kinetics}. This is reasonable considering the dynamics are only weakly perturbed (relative to the standard error) by the low excitation densities used here.

\begin{figure}[H]
    \centering
    \includegraphics[width=1\linewidth]{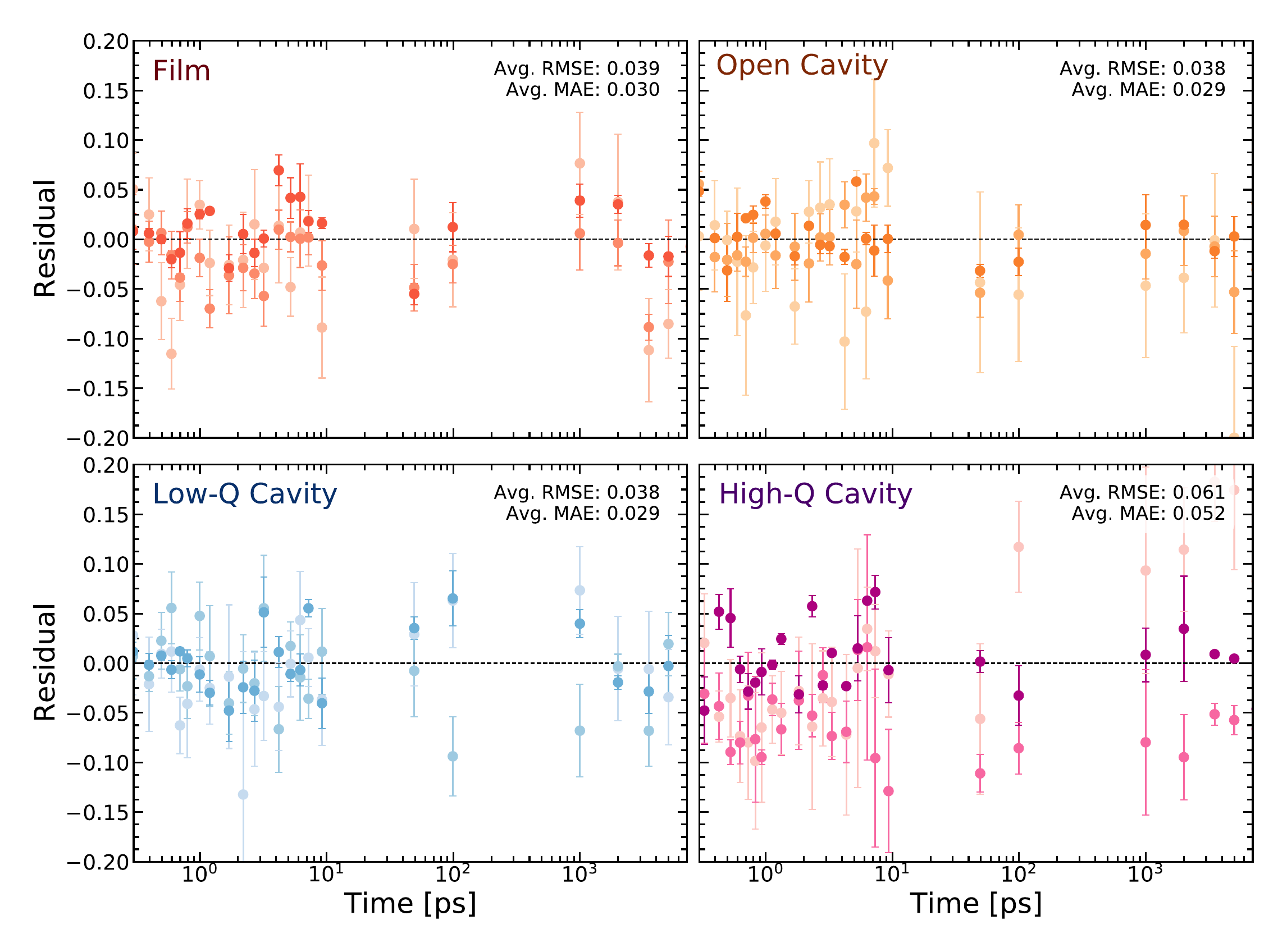}
       \caption{\textbf{Residuals and associated error of the fitted kinetics.} The residuals of each sample were calculated from the normalised kinetics shown in Figure \ref{SI_fig_kinetics}. For each sample, the root mean square error (RMSE) and the mean absolute error (MAE) were taken at individual pulse-energies, and then averaged across the three pulse energies studied in this chapter. In the fitting process, the average RMSE and average MAE were minimized for each sample, as shown in their respective panel.}
    \label{SI_fig_kinetics_residuals}
\end{figure}

To gauge how well the kinetic model fits the data, the residuals between the measured data points and model were calculated, as presented in Figure \ref{SI_fig_kinetics_residuals}.

Here we see the residuals lie randomly distributed above and below the zero-line, implying there is no systematic trend in the error. We find the high-Q cavity has the greatest RMSE and MAE, though this is likely due to the weaker signal-to-noise ratio stemming from a smaller excited-state population due to the high reflectivity of the top mirror.

\subsection{Detuning characterization}\label{SI_sec_detunings}

\begin{figure}[H]
    \centering
    \includegraphics[width=0.85\linewidth]{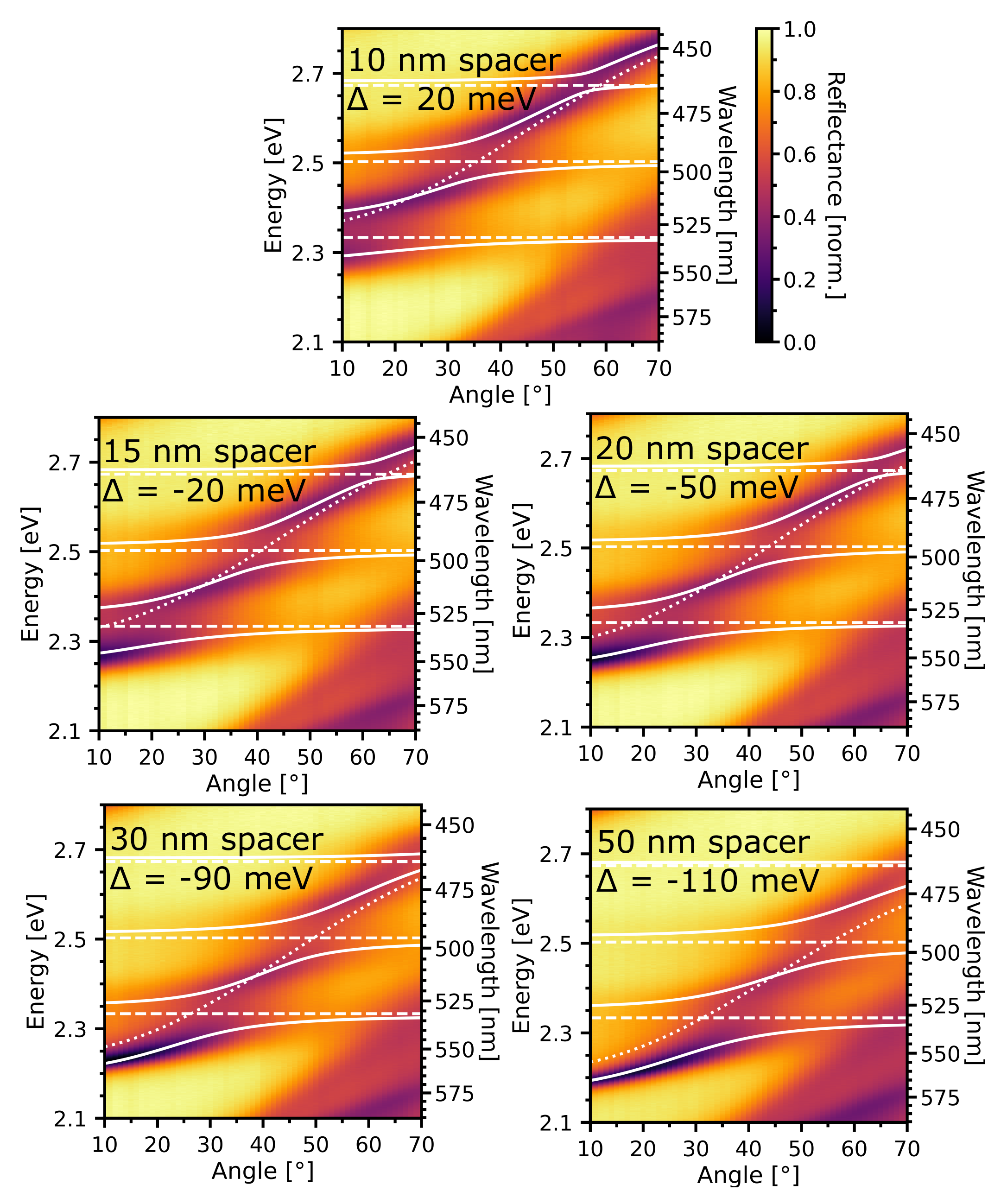}
    \caption{\textbf{Detuning dependence dispersions} Angle-resolved reflectance maps of the detuned cavities fabricated with different $\mathrm{SiO_2}$ spacer thicknesses, ranging from $\SI{10}{\nano\meter}$ to $\SI{50}{\nano\meter}$. The overlaid white lines show the coupled oscillator fitting results: the cavity mode (dotted line), the prominent rubrene $\mathrm{S_0 \to S_1}$ vibronic transitions (dashed lines), and the polariton branches (solid lines). The resulting detunings ($\Delta$) are indicated in each plot, where detuning is defined as the difference between the cavity mode energy at normal incidence and the $\mathrm{S_0 \to S_1}$ 0-0 transition energy.}
    \label{SI_fig_detuning_characterisation}
\end{figure}

\subsection{Supplementary EPR data}\label{SI_sec_EPR}

\begin{figure}[H]
    \centering
    \includegraphics[width=0.9\linewidth]{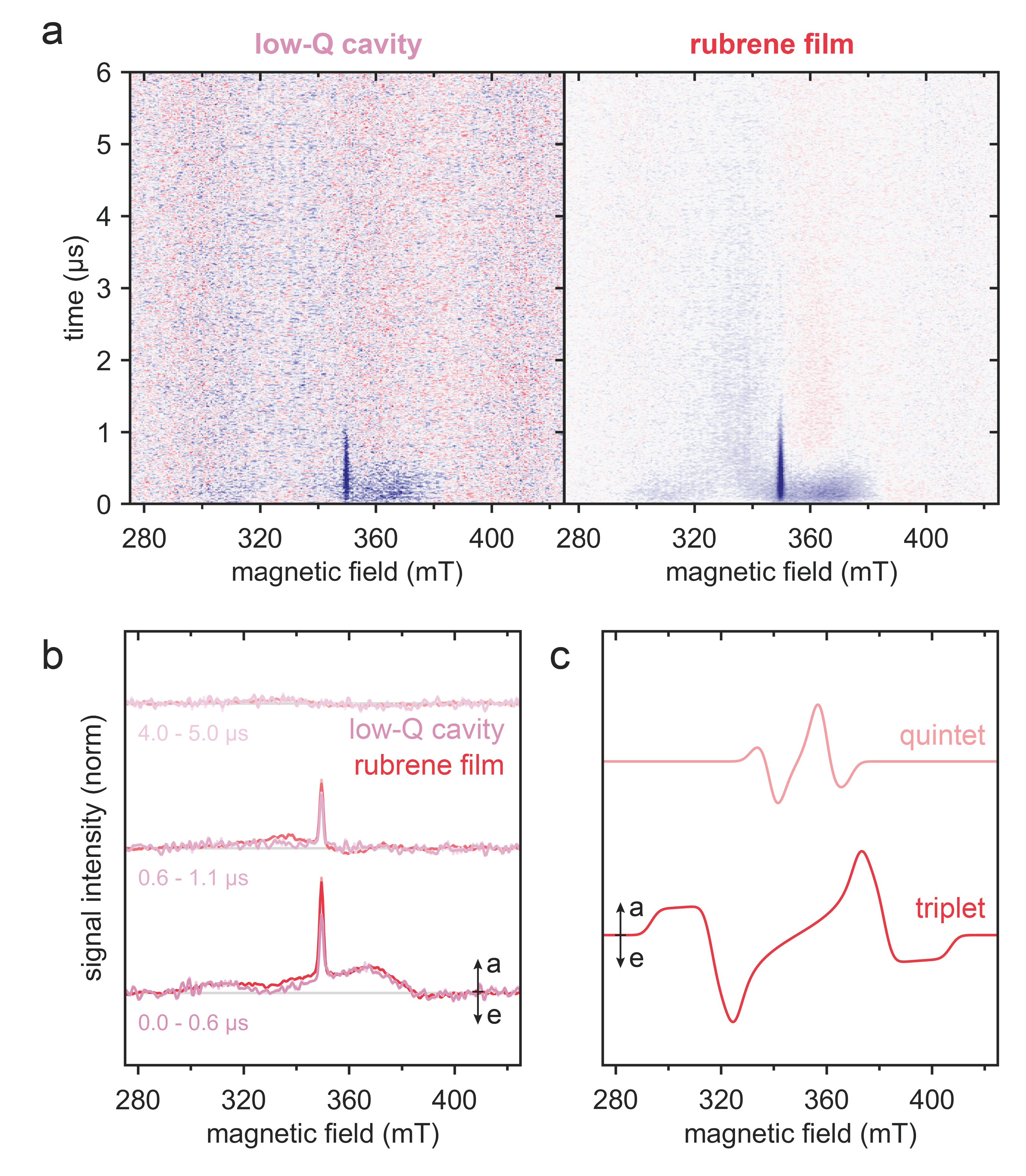}
       \caption{\textbf{Room-temperature transient EPR measurements of a rubrene film and the low-$Q$ cavity.} \textbf{(a)} Evolution of the EPR spectra as a function of time after laser excitation (blue = absorptive, red = emissive). \textbf{(b)} Normalized spectra extracted at different times after laser excitation. The low-Q cavity spectra (light-pink line) has been scaled to match the film's (red line) peak signal intensity. \textbf{(c)} Simulated EPR spectra assuming singlet fission only populates the $\mathrm{m}_S = 0$ spin sub-level for both the quintet and triplet states.}
    \label{fig_cavity_EPR}
\end{figure}

\begin{figure}[h]
    \centering
    \includegraphics[width=0.95\linewidth]{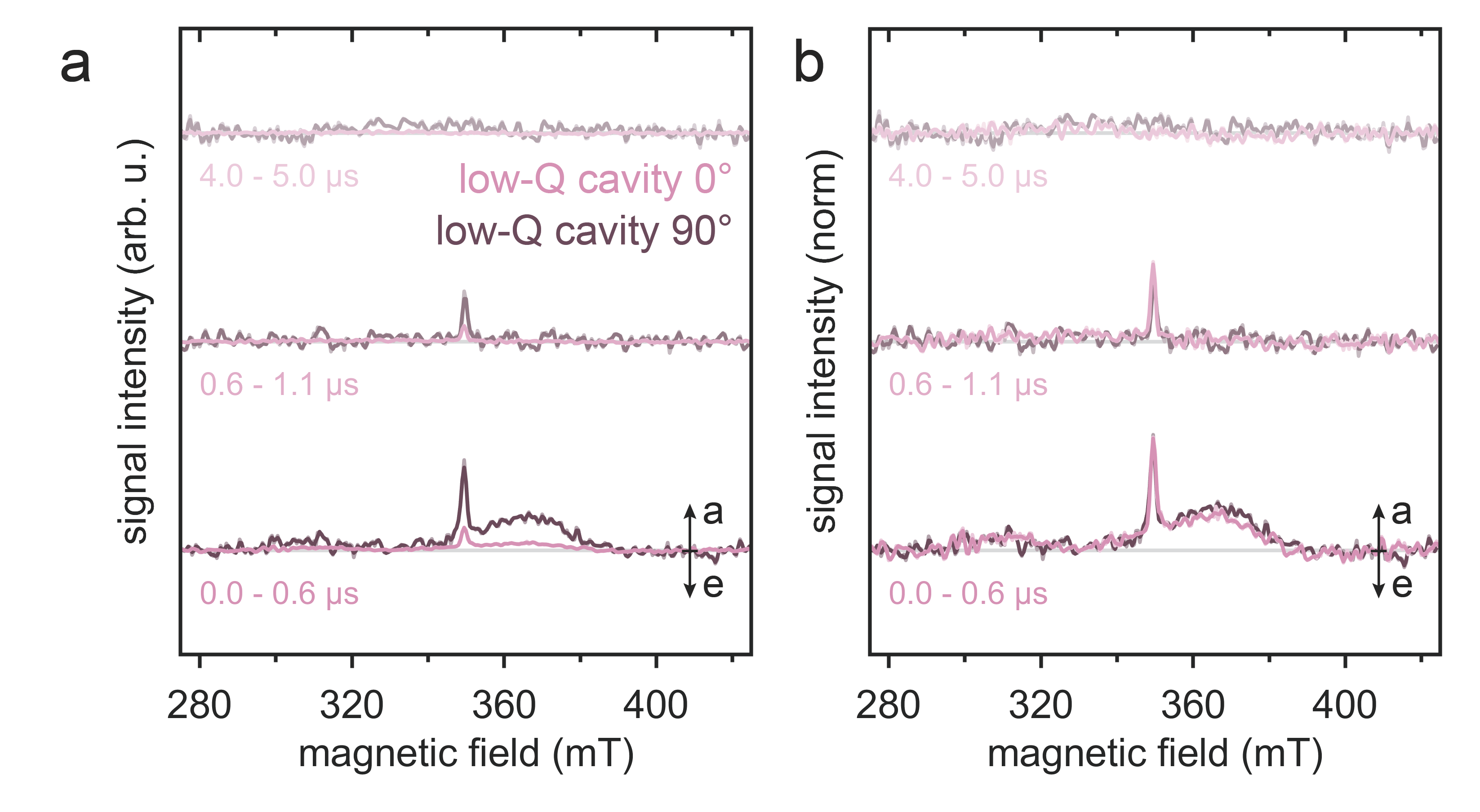}
       \caption{\textbf{Transient EPR spectra comparing excitation of the low-$Q$ cavity at normal incidence ($0^\circ$, light-pink) and on the edge of the substrate stack ($90^\circ$, purple).} \textbf{(a)} EPR spectra taken at the two different pump-sample orientations, extracted at different time windows after laser excitation. The signal intensity is enhanced at $90^\circ$, confirming the EPR signal of the cavity presented in Figure \ref{fig_cavity_EPR} is reduced due to the high-reflectance of the top DBR at the excitation wavelength. \textbf{(b)} Normalized EPR spectra at the two different pump-sample orientations. The spectra are identical, demonstrating the only changes are the signal intensity.}
    \label{C4_SI_fig_EPR_90deg}
\end{figure}

In Figure \ref{fig_cavity_EPR}a we compare the time-dependent polarization pattern ($\mathrm{ns}\text{--}\SI{}{\micro\second}$) of a rubrene film and strongly coupled cavity. The application of an external magnetic field lifts the degeneracy of the spin sub-levels, allowing transitions between these levels to be probed at a constant microwave frequency, resulting in either absorptive (positive) or emissive (negative) transition intensities due to a non-equilibrium population distribution across spin sublevels. Upon comparing the EPR spectra extracted at different times following photoexcitation, we find that the signal intensity associated with the cavity is significantly weaker than the bare film. This reduction is attributed to the high reflectivity of the distributed Bragg reflectors at the excitation wavelength ($\SI{505}{\nano\meter}$), which severely attenuates the pump intensity reaching the rubrene layer. This interpretation is confirmed by an auxiliary measurement performed at a $90^\circ$ orientation, where the laser beam was incident on the edge of the substrate stack, bypassing the $\mathrm{DBR}$ reflection; this yielded the same spectral shape but a significantly increased signal intensity (Figure \ref{C4_SI_fig_EPR_90deg}).

To examine the spectral features, we plot the normalized spectra at different time-delays, as shown in Figure \ref{fig_cavity_EPR}b. For reference, we have also simulated the EPR spectra expected for quintet-correlated triplet pair states as well as free triplet states with a spin polarization characteristic of $\mathrm{SF}$ (preferential population of the $\mathrm{m}_S = 0$ spin sub-level), as shown in Figure \ref{fig_cavity_EPR}c. The experimental room temperature spectra show no signs of quintet signatures and are instead in agreement with free triplet states formed by singlet fission.\cite{collinsFluctuatingExchangeInteractions2019} The observed triplet features are relatively broad, suggesting a large distribution of spin environments and therefore zero-field splitting parameters, which would be in agreement with the observed disorder in the rubrene layer.
Deviations from the spin polarization pattern expected in the presence of exclusive population of the $\mathrm{m}_S = 0$ sublevel, are likely due to the spin mixing mechanism and potentially influenced by triplet-triplet annihilation processes. Importantly though, the normalized spectra of the film and cavity overlap almost perfectly within the given signal-to-noise ratio across both early and late times after photoexcitation. The sharp and narrow absorptive signal located at approximately $\SI{350}{\milli\tesla}$ is likely due to a charge-transfer state or motionally averaged highly mobile triplet states.\cite{bualiIntersystemCrossingOutcompetes2025}

Apart from the signal intensity, the $\mathrm{EPR}$ spectra of the film and the low-$\mathrm{Q}$ cavity show no obvious differences; the triplet features remain virtually identical in shape and decay. This indicates that the dynamics of these triplet states and their spin-dependent formation are not perturbed by strong coupling.

\bibliography{library}

@article{berghuisEnhancedDelayedFluorescence2019,
  title = {Enhanced {{Delayed Fluorescence}} in {{Tetracene Crystals}} by {{Strong Light-Matter Coupling}}},
  author = {Berghuis, Anton Matthijs and Halpin, Alexei and Le-Van, Quynh and Ramezani, Mohammad and Wang, Shaojun and Murai, Shunsuke and Gómez Rivas, Jaime},
  year = {2019},
  journal = {Advanced Functional Materials},
  volume = {29},
  number = {36},
  pages = {1901317},
  issn = {1616-3028},
  doi = {10.1002/adfm.201901317},
  url = {https://onlinelibrary.wiley.com/doi/abs/10.1002/adfm.201901317},
  urlaccess = {2025-09-17},
  langid = {english}
}

@article{bossanyiOptimizedRubrenebasedNanoparticle2022,
  title = {In Optimized Rubrene-Based Nanoparticle Blends for Photon Upconversion, Singlet Energy Collection Outcompetes Triplet-Pair Separation, Not Singlet Fission},
  author = {Bossanyi, David G. and Sasaki, Yoichi and Wang, Shuanqing and Chekulaev, Dimitri and Kimizuka, Nobuo and Yanai, Nobuhiro and Clark, Jenny},
  year = {2022},
  journal = {Journal of Materials Chemistry C},
  shortjournal = {J. Mater. Chem. C},
  volume = {10},
  number = {12},
  pages = {4684--4696},
  publisher = {The Royal Society of Chemistry},
  issn = {2050-7534},
  doi = {10.1039/D1TC02955J},
  url = {https://pubs.rsc.org/en/content/articlelanding/2022/tc/d1tc02955j},
  urlaccess = {2025-09-17},
  langid = {english}
}

@article{bossanyiSingletFissionIncoherent2024,
  title = {Singlet Fission Is Incoherent in Pristine Orthorhombic Single Crystals of Rubrene: No Evidence of Triplet-Pair Emission},
  shorttitle = {Singlet Fission Is Incoherent in Pristine Orthorhombic Single Crystals of Rubrene},
  author = {Bossanyi, David G. and Matthiesen, Maik and Jayaprakash, Rahul and Bhattacharya, Sayantan and Zaumseil, Jana and Clark, Jenny},
  year = {2024},
  journal = {Faraday Discussions},
  shortjournal = {Faraday Discuss.},
  volume = {250},
  number = {0},
  pages = {162--180},
  publisher = {The Royal Society of Chemistry},
  issn = {1364-5498},
  doi = {10.1039/D3FD00150D},
  url = {https://pubs.rsc.org/en/content/articlelanding/2024/fd/d3fd00150d},
  urlaccess = {2025-09-17},
  langid = {english}
}

@article{climentNotDarkStrong2022,
  title = {Not Dark yet for Strong Light-Matter Coupling to Accelerate Singlet Fission Dynamics},
  author = {Climent, Clàudia and Casanova, David and Feist, Johannes and Garcia-Vidal, Francisco J.},
  year = {2022},
  journal = {Cell Reports Physical Science},
  shortjournal = {Cell Reports Physical Science},
  volume = {3},
  number = {4},
  pages = {100841},
  issn = {2666-3864},
  doi = {10.1016/j.xcrp.2022.100841},
  url = {https://www.sciencedirect.com/science/article/pii/S2666386422001151},
  urlaccess = {2025-09-17}
}

@article{colesVibrationallyAssistedPolaritonRelaxation2011,
  title = {Vibrationally {{Assisted Polariton-Relaxation Processes}} in {{Strongly Coupled Organic-Semiconductor Microcavities}}},
  author = {Coles, David M. and Michetti, Paolo and Clark, Caspar and Tsoi, Wing Chung and Adawi, Ali M. and Kim, Ji-Seon and Lidzey, David G.},
  year = {2011},
  journal = {Advanced Functional Materials},
  volume = {21},
  number = {19},
  pages = {3691--3696},
  issn = {1616-3028},
  doi = {10.1002/adfm.201100756},
  url = {https://onlinelibrary.wiley.com/doi/abs/10.1002/adfm.201100756},
  urlaccess = {2025-09-17},
  langid = {english}
}

@article{collinsFluctuatingExchangeInteractions2019,
  title = {Fluctuating Exchange Interactions Enable Quintet Multiexciton Formation in Singlet Fission},
  author = {Collins, Miles I. and McCamey, Dane R. and Tayebjee, Murad J. Y.},
  year = {2019},
  journal = {The Journal of Chemical Physics},
  shortjournal = {J. Chem. Phys.},
  volume = {151},
  number = {16},
  pages = {164104},
  issn = {0021-9606},
  doi = {10.1063/1.5115816},
  url = {https://doi.org/10.1063/1.5115816},
  urlaccess = {2025-09-17}
}

@article{dayImprovingSpectralModification2019,
  title = {Improving Spectral Modification for Applications in Solar Cells: {{A}} Review},
  author = {Day, Joseph and Senthilarasu, S. and Mallick, Tapas K.},
  year = {2019},
  journal = {Renewable Energy},
  volume = {132},
  pages = {186--205},
  publisher = {Elsevier Ltd},
  issn = {18790682},
  doi = {10.1016/j.renene.2018.07.101}
}

@article{eiznerInvertingSingletTriplet2019,
  title = {Inverting Singlet and Triplet Excited States Using Strong Light-Matter Coupling},
  author = {Eizner, Elad and Martínez-Martínez, Luis A. and Yuen-Zhou, Joel and Kéna-Cohen, Stéphane},
  year = {2019},
  journal = {Science Advances},
  volume = {5},
  number = {12},
  eprint = {31840063},
  eprinttype = {pubmed},
  pages = {1--9},
  issn = {23752548},
  doi = {10.1126/sciadv.aax4482}
}

@article{grantEfficientRadiativePumping2016,
  title = {Efficient {{Radiative Pumping}} of {{Polaritons}} in a {{Strongly Coupled Microcavity}} by a {{Fluorescent Molecular Dye}}},
  author = {Grant, Richard T. and Michetti, Paolo and Musser, Andrew J. and Gregoire, Pascal and Virgili, Tersilla and Vella, Eleonora and Cavazzini, Marco and Georgiou, Kyriacos and Galeotti, Francesco and Clark, Caspar and Clark, Jenny and Silva, Carlos and Lidzey, David G.},
  year = {2016},
  journal = {Advanced Optical Materials},
  volume = {4},
  number = {10},
  pages = {1615--1623},
  issn = {21951071},
  doi = {10.1002/adom.201600337}
}

@article{guOpticalCavityManipulationConical2021,
  title = {Optical-{{Cavity Manipulation}} of {{Conical Intersections}} and {{Singlet Fission}} in {{Pentacene Dimers}}},
  author = {Gu, Bing and Mukamel, Shaul},
  year = {2021},
  journal = {Journal of Physical Chemistry Letters},
  volume = {12},
  number = {8},
  eprint = {33615792},
  eprinttype = {pubmed},
  pages = {2052--2056},
  publisher = {American Chemical Society},
  issn = {19487185},
  doi = {10.1021/acs.jpclett.0c03829}
}

@article{haegelPhotovoltaicsMultiterawattScale2023,
  title = {Photovoltaics at Multi-Terawatt Scale: {{Waiting}} Is Not an Option},
  shorttitle = {Photovoltaics at Multi-Terawatt Scale},
  author = {Haegel, Nancy M. and Verlinden, Pierre and Victoria, Marta and Altermatt, Pietro and Atwater, Harry and Barnes, Teresa and Breyer, Christian and Case, Chris and De Wolf, Stefaan and Deline, Chris and Dharmrin, Marwan and Dimmler, Bernhard and Gloeckler, Markus and Goldschmidt, Jan Christoph and Hallam, Brett and Haussener, Sophia and Holder, Burkhard and Jaeger, Ulrich and Jaeger-Waldau, Arnulf and Kaizuka, Izumi and Kikusato, Hiroshi and Kroposki, Benjamin and Kurtz, Sarah and Matsubara, Koji and Nowak, Stefan and Ogimoto, Kazuhiko and Peter, Christian and Peters, Ian Marius and Philipps, Simon and Powalla, Michael and Rau, Uwe and Reindl, Thomas and Roumpani, Maria and Sakurai, Keiichiro and Schorn, Christian and Schossig, Peter and Schlatmann, Rutger and Sinton, Ron and Slaoui, Abdelilah and Smith, Brittany L. and Schneidewind, Peter and family=Stanbery, given=BJ, given-i=BJ and Topic, Marko and Tumas, William and Vasi, Juzer and Vetter, Matthias and Weber, Eicke and Weeber, A. W. and Weidlich, Anke and Weiss, Dirk and Bett, Andreas W.},
  year = {2023},
  journal = {Science},
  volume = {380},
  number = {6640},
  pages = {39--42},
  publisher = {American Association for the Advancement of Science},
  doi = {10.1126/science.adf6957},
  url = {https://www.science.org/doi/full/10.1126/science.adf6957},
  urlaccess = {2025-09-17}
}

@article{Herrera2016,
  title = {Cavity-{{Controlled Chemistry}} in {{Molecular Ensembles}}},
  author = {Herrera, Felipe and Spano, Frank C.},
  year = {2016},
  journal = {Physical Review Letters},
  volume = {116},
  number = {23},
  eprint = {1512.05017},
  eprinttype = {arXiv},
  pages = {1--6},
  issn = {10797114},
  doi = {10.1103/PhysRevLett.116.238301}
}

@article{hobsonStrongExcitonPhoton2002,
  title = {Strong Exciton–Photon Coupling in a Low-{{Q}} All-Metal Mirror Microcavity},
  author = {Hobson, P. A. and Barnes, W. L. and Lidzey, D. G. and Gehring, G. A. and Whittaker, D. M. and Skolnick, M. S. and Walker, S.},
  year = {2002},
  journal = {Applied Physics Letters},
  shortjournal = {Appl. Phys. Lett.},
  volume = {81},
  number = {19},
  pages = {3519--3521},
  issn = {0003-6951},
  doi = {10.1063/1.1517714},
  url = {https://doi.org/10.1063/1.1517714},
  urlaccess = {2025-09-17}
}

@article{Hopfield1958,
  title = {Theory of the Contribution of Excitons to the Complex Dielectric Constant of Crystals},
  author = {Hopfield, J. J.},
  year = {1958},
  journal = {Physical Review},
  volume = {112},
  number = {5},
  pages = {1555--1567},
  issn = {0031899X},
  doi = {10.1103/PhysRev.112.1555}
}

@article{imperatoreReproducibilityCavityenhancedChemical2021a,
  title = {Reproducibility of Cavity-Enhanced Chemical Reaction Rates in the Vibrational Strong Coupling Regime},
  author = {Imperatore, Mario V. and Asbury, John B. and Giebink, Noel C.},
  year = {2021},
  journal = {The Journal of Chemical Physics},
  shortjournal = {J. Chem. Phys.},
  volume = {154},
  number = {19},
  pages = {191103},
  issn = {0021-9606},
  doi = {10.1063/5.0046307},
  url = {https://doi.org/10.1063/5.0046307},
  urlaccess = {2025-09-17}
}

@article{liuRoleLongLivedExcitons2020,
  title = {The {{Role}} of {{Long-Lived Excitons}} in the {{Dynamics}} of {{Strongly Coupled Molecular Polaritons}}},
  author = {Liu, Bin and Menon, Vinod M. and Sfeir, Matthew Y.},
  year = {2020},
  journal = {ACS Photonics},
  volume = {7},
  number = {8},
  eprint = {2002.09747},
  eprinttype = {arXiv},
  pages = {2292--2301},
  publisher = {American Chemical Society},
  issn = {23304022},
  doi = {10.1021/acsphotonics.0c00895}
}

@article{martinez-martinezPolaritonAssistedSingletFission2018,
  title = {Polariton-{{Assisted Singlet Fission}} in {{Acene Aggregates}}},
  author = {Martínez-Martínez, Luis A. and Du, Matthew and Ribeiro, Raphael F. and Kéna-Cohen, Stéphane and Yuen-Zhou, Joel},
  year = {2018},
  journal = {Journal of Physical Chemistry Letters},
  volume = {9},
  number = {8},
  eprint = {29551074},
  eprinttype = {pubmed},
  pages = {1951--1957},
  issn = {19487185},
  doi = {10.1021/acs.jpclett.8b00008}
}

@article{maslennikovInterplayMixedPure2025,
  title = {Interplay between {{Mixed}} and {{Pure Exciton States Controls Singlet Fission}} in {{Rubrene Single Crystals}}},
  author = {Maslennikov, Dmitry R. and Maimaris, Marios and Ning, Haoqing and Zheng, Xijia and Mondal, Navendu and Bruevich, Vladimir V. and Pratik, Saied Md and Dong, Yifan and Tisch, John W. G. and Musser, Andrew J. and Podzorov, Vitaly and Bredas, Jean-Luc and Coropceanu, Veaceslav and Bakulin, Artem A.},
  year = {2025},
  journal = {Journal of the American Chemical Society},
  shortjournal = {J. Am. Chem. Soc.},
  volume = {147},
  number = {27},
  pages = {23536--23544},
  publisher = {American Chemical Society},
  issn = {0002-7863},
  doi = {10.1021/jacs.5c02993},
  url = {https://doi.org/10.1021/jacs.5c02993},
  urlaccess = {2025-09-17}
}

@article{michailAddressingDarkState2024,
  title = {Addressing the {{Dark State Problem}} in {{Strongly Coupled Organic Exciton-Polariton Systems}}},
  author = {Michail, Evripidis and Rashidi, Kamyar and Liu, Bin and He, Guiying and Menon, Vinod M. and Sfeir, Matthew Y.},
  year = {2024},
  journal = {Nano Letters},
  shortjournal = {Nano Lett.},
  volume = {24},
  number = {2},
  pages = {557--565},
  publisher = {American Chemical Society},
  issn = {1530-6984},
  doi = {10.1021/acs.nanolett.3c02984},
  url = {https://doi.org/10.1021/acs.nanolett.3c02984},
  urlaccess = {2025-09-17}
}

@article{Miyata2017,
  title = {Coherent Singlet Fission Activated by Symmetry Breaking},
  author = {Miyata, Kiyoshi and Kurashige, Yuki and Watanabe, Kazuya and Sugimoto, Toshiki and Takahashi, Shota and Tanaka, Shunsuke and Takeya, Jun and Yanai, Takeshi and Matsumoto, Yoshiyasu},
  year = {2017},
  journal = {Nature Chemistry},
  volume = {9},
  number = {10},
  eprint = {28937675},
  eprinttype = {pubmed},
  pages = {983--989},
  publisher = {Nature Publishing Group},
  issn = {17554349},
  doi = {10.1038/nchem.2784}
}

@article{monyAngleIndependentPolaritonEmission2018,
  title = {Angle-{{Independent Polariton Emission Lifetime Shown}} by {{Perylene Hybridized}} to the {{Vacuum Field Inside}} a {{Fabry}}–{{Pérot Cavity}}},
  author = {Mony, Jürgen and Hertzog, Manuel and Kushwaha, Khushbu and Börjesson, Karl},
  year = {2018},
  journal = {The Journal of Physical Chemistry C},
  shortjournal = {J. Phys. Chem. C},
  volume = {122},
  number = {43},
  pages = {24917--24923},
  publisher = {American Chemical Society},
  issn = {1932-7447},
  doi = {10.1021/acs.jpcc.8b07283},
  url = {https://doi.org/10.1021/acs.jpcc.8b07283},
  urlaccess = {2025-09-17}
}

@article{musserTripletPairStatesOrganic2019,
  title = {Triplet-{{Pair States}} in {{Organic Semiconductors}}},
  author = {Musser, Andrew J. and Clark, Jenny},
  year = {2019},
  journal = {Annual Review of Physical Chemistry},
  volume = {70},
  eprint = {31174458},
  eprinttype = {pubmed},
  pages = {323--351},
  issn = {15451593},
  doi = {10.1146/annurev-physchem-042018-052435}
}

@article{renkenUntargetedEffectsOrganic2021,
  title = {Untargeted Effects in Organic Exciton–Polariton Transient Spectroscopy: {{A}} Cautionary Tale},
  shorttitle = {Untargeted Effects in Organic Exciton–Polariton Transient Spectroscopy},
  author = {Renken, Scott and Pandya, Raj and Georgiou, Kyriacos and Jayaprakash, Rahul and Gai, Lizhi and Shen, Zhen and Lidzey, David G. and Rao, Akshay and Musser, Andrew J.},
  year = {2021},
  journal = {The Journal of Chemical Physics},
  shortjournal = {J. Chem. Phys.},
  volume = {155},
  number = {15},
  pages = {154701},
  issn = {0021-9606},
  doi = {10.1063/5.0063173},
  url = {https://doi.org/10.1063/5.0063173},
  urlaccess = {2025-09-17}
}

@article{riderSomethingNothingLinking2021,
  title = {Something from Nothing: Linking Molecules with Virtual Light},
  shorttitle = {Something from Nothing},
  author = {Rider, M. S. and Barnes, W. L.},
  year = {2021},
  journal = {Contemporary Physics},
  volume = {62},
  number = {4},
  pages = {217--232},
  publisher = {Taylor \& Francis},
  issn = {0010-7514},
  doi = {10.1080/00107514.2022.2101749},
  url = {https://doi.org/10.1080/00107514.2022.2101749},
  urlaccess = {2025-09-17}
}

@article{schwennickeWhenMolecularPolaritons2025,
  title = {When Do Molecular Polaritons Behave like Optical Filters?},
  author = {Schwennicke, Kai and Koner, Arghadip and Pérez-Sánchez, Juan B. and Xiong, Wei and Giebink, Noel C. and Weichman, Marissa L. and Yuen-Zhou, Joel},
  year = {2025},
  journal = {Chemical Society Reviews},
  shortjournal = {Chem. Soc. Rev.},
  volume = {54},
  number = {13},
  pages = {6482--6504},
  publisher = {The Royal Society of Chemistry},
  issn = {1460-4744},
  doi = {10.1039/D4CS01024H},
  url = {https://pubs.rsc.org/en/content/articlelanding/2025/cs/d4cs01024h},
  urlaccess = {2025-09-17},
  langid = {english}
}

@article{shockleyDetailedBalanceLimit1961,
  title = {Detailed {{Balance Limit}} of {{Efficiency}} of P‐n {{Junction Solar Cells}}},
  author = {Shockley, William and Queisser, Hans J.},
  year = {1961},
  journal = {Journal of Applied Physics},
  shortjournal = {J. Appl. Phys.},
  volume = {32},
  number = {3},
  pages = {510--519},
  issn = {0021-8979},
  doi = {10.1063/1.1736034},
  url = {https://doi.org/10.1063/1.1736034},
  urlaccess = {2025-09-17}
}

@article{takahashiSingletFissionAmorphous2019,
  title = {Singlet Fission of Amorphous Rubrene Modulated by Polariton Formation},
  author = {Takahashi, Shota and Watanabe, Kazuya and Matsumoto, Yoshiyasu},
  year = {2019},
  journal = {The Journal of Chemical Physics},
  shortjournal = {J. Chem. Phys.},
  volume = {151},
  number = {7},
  pages = {074703},
  issn = {0021-9606},
  doi = {10.1063/1.5108698},
  url = {https://doi.org/10.1063/1.5108698},
  urlaccess = {2025-09-17}
}

@article{tayebjeeShockleyqueisserMolecularApproaches2015,
  title = {Beyond Shockley-Queisser: {{Molecular}} Approaches to High-Efficiency Photovoltaics},
  author = {Tayebjee, Murad J.Y. and McCamey, Dane R. and Schmidt, Timothy W.},
  year = {2015},
  journal = {Journal of Physical Chemistry Letters},
  volume = {6},
  number = {12},
  pages = {2367--2378},
  publisher = {American Chemical Society},
  issn = {19487185},
  doi = {10.1021/acs.jpclett.5b00716}
}

@article{thomasNonPolaritonicEffectsCavityModified2024b,
  title = {Non-{{Polaritonic Effects}} in {{Cavity-Modified Photochemistry}}},
  author = {Thomas, Philip A. and Tan, Wai Jue and Kravets, Vasyl G. and Grigorenko, Alexander N. and Barnes, William L.},
  year = {2024},
  journal = {Advanced Materials},
  volume = {36},
  number = {7},
  pages = {2309393},
  issn = {1521-4095},
  doi = {10.1002/adma.202309393},
  url = {https://onlinelibrary.wiley.com/doi/abs/10.1002/adma.202309393},
  urlaccess = {2025-09-17},
  langid = {english}
}

@article{thomasSelectionBiasStrong2024,
  title = {Selection {{Bias}} in {{Strong Coupling Experiments}}},
  author = {Thomas, Philip A. and Barnes, William L.},
  year = {2024},
  journal = {Journal of Physical Chemistry Letters},
  volume = {15},
  number = {6},
  eprint = {38356453},
  eprinttype = {pubmed},
  pages = {1708--1710},
  publisher = {American Chemical Society},
  issn = {19487185},
  doi = {10.1021/acs.jpclett.3c03546}
}

@article{thompsonObservationNormalmodeSplitting1992,
  title = {Observation of Normal-Mode Splitting for an Atom in an Optical Cavity},
  author = {Thompson, R. J. and Rempe, G. and Kimble, H. J.},
  year = {1992},
  journal = {Physical Review Letters},
  shortjournal = {Phys. Rev. Lett.},
  volume = {68},
  number = {8},
  pages = {1132--1135},
  issn = {0031-9007},
  doi = {10.1103/PhysRevLett.68.1132},
  url = {https://link.aps.org/doi/10.1103/PhysRevLett.68.1132},
  urlaccess = {2025-09-17},
  langid = {english}
}

@article{volekStructuralDisorderEdges2023,
  title = {Structural {{Disorder}} at the {{Edges}} of {{Rubrene Crystals Enhances Singlet Fission}}},
  author = {Volek, Tanner S. and Armstrong, Zachary T. and Sowa, Jakub K. and Wilson, Kelly S. and Bohlmann Kunz, Miriam and Bera, Kajari and Koble, Ma Kenna and Frontiera, Renee R. and Rossky, Peter J. and Zanni, Martin T. and Roberts, Sean T.},
  year = {2023},
  journal = {Journal of Physical Chemistry Letters},
  volume = {14},
  number = {50},
  eprint = {38088867},
  eprinttype = {pubmed},
  pages = {11497--11505},
  publisher = {American Chemical Society},
  issn = {19487185},
  doi = {10.1021/acs.jpclett.3c02845}
}

@article{weisbuchObservationCoupledExcitonphoton1992,
  title = {Observation of the Coupled Exciton-Photon Mode Splitting in a Semiconductor Quantum Microcavity},
  author = {Weisbuch, C. and Nishioka, M. and Ishikawa, A. and Arakawa, Y.},
  year = {1992},
  journal = {Physical Review Letters},
  shortjournal = {Phys. Rev. Lett.},
  volume = {69},
  number = {23},
  pages = {3314--3317},
  issn = {0031-9007},
  doi = {10.1103/PhysRevLett.69.3314},
  url = {https://link.aps.org/doi/10.1103/PhysRevLett.69.3314},
  urlaccess = {2025-09-17},
  langid = {english}
}

@article{Yago2016,
  title = {Magnetic Field Effects on Triplet Pair Generated by Singlet Fission in an Organic Crystal: {{Application}} of Radical Pair Model to Triplet Pair},
  author = {Yago, Tomoaki and Ishikawa, Kei and Katoh, Ryuzi and Wakasa, Masanobu},
  year = {2016},
  journal = {Journal of Physical Chemistry C},
  volume = {120},
  number = {49},
  pages = {27858--27870},
  issn = {19327455},
  doi = {10.1021/acs.jpcc.6b09570}
}

@article{yoonUltrasmoothOrganicFilms2021,
  title = {Ultrasmooth {{Organic Films Via Efficient Aggregation Suppression}} by a {{Low-Vacuum Physical Vapor Deposition}}},
  author = {Yoon, Youngkwan and Lee, Jinho and Lee, Seulgi and Kim, Soyoung and Choi, Hee Cheul},
  year = {2021},
  journal = {Materials},
  volume = {14},
  number = {23},
  pages = {7247},
  publisher = {Multidisciplinary Digital Publishing Institute},
  issn = {1996-1944},
  doi = {10.3390/ma14237247},
  url = {https://www.mdpi.com/1996-1944/14/23/7247},
  urlaccess = {2025-09-17},
  langid = {english}
}

@article{zhangJointEffectsExcitonExciton2021,
  title = {Joint {{Effects}} of {{Exciton}}-{{Exciton}} and {{Exciton}}-{{Photon Couplings}} on the {{Singlet Fission Dynamics}} in {{Organic Aggregates}}},
  author = {Zhang, Bin and Zhao, Yi and Liang, Wan Zhen},
  year = {2021},
  journal = {Journal of Physical Chemistry C},
  volume = {125},
  number = {3},
  pages = {1654--1664},
  publisher = {American Chemical Society},
  issn = {19327455},
  doi = {10.1021/ACS.JPCC.0C08185}
}

@article{leeLightEmittingRubreneNanowire2009,
  title = {Light-{{Emitting Rubrene Nanowire Arrays}}: {{A Comparison}} with {{Rubrene Single Crystals}}},
  shorttitle = {Light-{{Emitting Rubrene Nanowire Arrays}}},
  author = {Lee, Jin Woo and Kim, Kihyun and Park, Dong Hyuk and Cho, Mi Yeon and Lee, Yong Baek and Jung, Jin Sun and Kim, Dae-Chul and Kim, Jeongyong and Joo, Jinsoo},
  year = {2009},
  journal = {Advanced Functional Materials},
  volume = {19},
  number = {5},
  pages = {704--710},
  issn = {1616-3028},
  doi = {10.1002/adfm.200801180},
  url = {https://onlinelibrary.wiley.com/doi/abs/10.1002/adfm.200801180},
  urlaccess = {2025-09-19},
  langid = {english}
}

@article{spanoSpectralSignaturesFrenkel2010,
  title = {The {{Spectral Signatures}} of {{Frenkel Polarons}} in {{H-}} and {{J-Aggregates}}},
  author = {Spano, Frank C.},
  year = {2010},
  journal = {Accounts of Chemical Research},
  shortjournal = {Acc. Chem. Res.},
  volume = {43},
  number = {3},
  pages = {429--439},
  publisher = {American Chemical Society},
  issn = {0001-4842},
  doi = {10.1021/ar900233v},
  url = {https://doi.org/10.1021/ar900233v},
  urlaccess = {2025-09-19}
}

@article{bergantinPressureInducedConformationalChange2014,
  title = {Pressure-{{Induced Conformational Change}} in {{Organic Semiconductors}}: {{Triggering}} a {{Reversible Phase Transition}} in {{Rubrene}}},
  shorttitle = {Pressure-{{Induced Conformational Change}} in {{Organic Semiconductors}}},
  author = {Bergantin, Stefano and Moret, Massimo and Buth, Gernot and Fabbiani, Francesca P. A.},
  year = {2014},
  journal = {The Journal of Physical Chemistry C},
  shortjournal = {J. Phys. Chem. C},
  volume = {118},
  number = {25},
  pages = {13476--13483},
  publisher = {American Chemical Society},
  issn = {1932-7447},
  doi = {10.1021/jp503271h},
  url = {https://doi.org/10.1021/jp503271h},
  urlaccess = {2025-09-19}
}

@article{beraFemtosecondRamanMicroscopy2017,
  title = {Femtosecond {{Raman Microscopy Reveals Structural Dynamics Leading}} to {{Triplet Separation}} in {{Rubrene Singlet Fission}}},
  author = {Bera, Kajari and Douglas, Christopher J. and Frontiera, Renee R.},
  year = {2017},
  journal = {The Journal of Physical Chemistry Letters},
  shortjournal = {J. Phys. Chem. Lett.},
  volume = {8},
  number = {23},
  pages = {5929--5934},
  publisher = {American Chemical Society},
  doi = {10.1021/acs.jpclett.7b02769},
  url = {https://doi.org/10.1021/acs.jpclett.7b02769},
  urlaccess = {2025-09-24}
}

@article{breenTripletSeparationDrives2017,
  title = {Triplet {{Separation Drives Singlet Fission}} after {{Femtosecond Correlated Triplet Pair Production}} in {{Rubrene}}},
  author = {Breen, Ilana and Tempelaar, Roel and Bizimana, Laurie A. and Kloss, Benedikt and Reichman, David R. and Turner, Daniel B.},
  year = {2017},
  journal = {Journal of the American Chemical Society},
  shortjournal = {J. Am. Chem. Soc.},
  volume = {139},
  number = {34},
  pages = {11745--11751},
  publisher = {American Chemical Society},
  issn = {0002-7863},
  doi = {10.1021/jacs.7b02621},
  url = {https://doi.org/10.1021/jacs.7b02621},
  urlaccess = {2025-09-24}
}

@article{ostroverkhova_excitons_2024,
	title = {{Excitons and Polaritons in Singlet Fission Materials: Photophysics, Photochemistry, and Optoelectronics}},
	author = {Ostroverkhova, Oksana and Goldthwaite, Winston and Lamug, Roshell},
	year = {2024},
	journal = {{MRS} Bulletin},
	volume = {49},
	number = {9},
	pages = {873--884},
	doi = {10.1557/s43577-024-00761-2},
	url = {https://doi.org/10.1557/s43577-024-00761-2},
}

@article{herkstroeterTripletStateEnergies1981,
  title = {The Triplet State Energies of Rubrene and Diphenylisobenzofuran},
  author = {Herkstroeter, W. G. and Merkel, P. B.},
  year = {1981},
  journal = {Journal of Photochemistry},
  shortjournal = {Journal of Photochemistry},
  volume = {16},
  number = {3},
  pages = {331--342},
  issn = {0047-2670},
  doi = {10.1016/0047-2670(81)80043-3},
  url = {https://www.sciencedirect.com/science/article/pii/0047267081800433},
  urlaccess = {2025-10-06}
}

@article{singh-rachfordPdIIPhthalocyanineSensitizedTripletTriplet2008,
  title = {Pd({{II}}) {{Phthalocyanine-Sensitized Triplet}}-{{Triplet Annihilation}} from {{Rubrene}}},
  author = {Singh-Rachford, Tanya N. and Castellano, Felix N.},
  year = {2008},
  journal = {The Journal of Physical Chemistry A},
  shortjournal = {J. Phys. Chem. A},
  volume = {112},
  number = {16},
  pages = {3550--3556},
  publisher = {American Chemical Society},
  issn = {1089-5639},
  doi = {10.1021/jp7111878},
  url = {https://doi.org/10.1021/jp7111878},
  urlaccess = {2025-10-06}
}

@article{bossanyiSpinStatisticsTriplet2021,
  title = {Spin {{Statistics}} for {{Triplet}}–{{Triplet Annihilation Upconversion}}: {{Exchange Coupling}}, {{Intermolecular Orientation}}, and {{Reverse Intersystem Crossing}}},
  shorttitle = {Spin {{Statistics}} for {{Triplet}}–{{Triplet Annihilation Upconversion}}},
  author = {Bossanyi, David G. and Sasaki, Yoichi and Wang, Shuangqing and Chekulaev, Dimitri and Kimizuka, Nobuo and Yanai, Nobuhiro and Clark, Jenny},
  year = {2021},
  journal = {JACS Au},
  shortjournal = {JACS Au},
  volume = {1},
  number = {12},
  pages = {2188--2201},
  publisher = {American Chemical Society},
  doi = {10.1021/jacsau.1c00322},
  url = {https://doi.org/10.1021/jacsau.1c00322},
  urlaccess = {2025-10-06}
}

@article{maSingletFissionRubrene2012a,
  title = {Singlet Fission in Rubrene Single Crystal: Direct Observation by Femtosecond Pump–Probe Spectroscopy},
  shorttitle = {Singlet Fission in Rubrene Single Crystal},
  author = {Ma, Lin and Zhang, Keke and Kloc, Christian and Sun, Handong and Michel-Beyerle, Maria E. and Gurzadyan, Gagik G.},
  year = {2012},
  journal = {Physical Chemistry Chemical Physics},
  shortjournal = {Phys. Chem. Chem. Phys.},
  volume = {14},
  number = {23},
  pages = {8307--8312},
  publisher = {The Royal Society of Chemistry},
  issn = {1463-9084},
  doi = {10.1039/C2CP40449D},
  url = {https://pubs.rsc.org/en/content/articlelanding/2012/cp/c2cp40449d},
  urlaccess = {2025-10-06},
  langid = {english}
}

@article{jaumotMCRALSGUI202015,
  title = {{MCR-ALS GUI} 2.0: New Features and Applications},
  author = {Jaumot, Joaquim and de Juan, Anna and Tauler, Rom{\`a}},
  year = {2015},
  journal = {Chemometrics and Intelligent Laboratory Systems},
  volume = {140},
  pages = {1--12},
  doi = {10.1016/j.chemolab.2014.10.003}
}

@article{pilandMagneticFieldEffects2013b,
  title = {Magnetic {{Field Effects}} on {{Singlet Fission}} and {{Fluorescence Decay Dynamics}} in {{Amorphous Rubrene}}},
  author = {Piland, Geoffrey B. and Burdett, Jonathan J. and Kurunthu, Dharmalingam and Bardeen, Christopher J.},
  year = {2013},
  journal = {The Journal of Physical Chemistry C},
  shortjournal = {J. Phys. Chem. C},
  volume = {117},
  number = {3},
  pages = {1224--1236},
  publisher = {American Chemical Society},
  issn = {1932-7447},
  doi = {10.1021/jp309286v},
  url = {https://doi.org/10.1021/jp309286v},
  urlaccess = {2025-10-06}
}

@article{litinskayaFastPolaritonRelaxation2004a,
  title = {Fast Polariton Relaxation in Strongly Coupled Organic Microcavities},
  author = {Litinskaya, M. and Reineker, P. and Agranovich, V. M.},
  year = {2004},
  journal = {Journal of Luminescence},
  shortjournal = {Journal of Luminescence},
  series = {325th {{Wilhelm}} and {{Else Heraeus Workshop}}. {{Organic Molecular Solids}} : {{Excited Electronic States}} and {{Optical Properties}}},
  volume = {110},
  number = {4},
  pages = {364--372},
  issn = {0022-2313},
  doi = {10.1016/j.jlumin.2004.08.033},
  url = {https://www.sciencedirect.com/science/article/pii/S002223130400273X},
  urlaccess = {2025-10-09}
}

@article{wang_quantum_2014,
  title = {Quantum Yield of Polariton Emission from Hybrid Light-Matter States},
  author = {Wang, Shaojun and Chervy, Thibault and George, Jino and Hutchison, James A. and Genet, Cyriaque and Ebbesen, Thomas W.},
  journal = {The Journal of Physical Chemistry Letters},
  volume = {5},
  number = {8},
  pages = {1433--1439},
  year = {2014},
  month = {4},
  doi = {10.1021/jz5004439},
  eprint = {https://pubs.acs.org/doi/10.1021/jz5004439},
  eprinttype = {url},
  issn = {1948-7185},
  language = {en},
}

@article{leSlowSingletFission2016,
  title = {Slow {{Singlet Fission Observed}} in a {{Polycrystalline Perylenediimide Thin Film}}},
  author = {Le, Aaron K. and Bender, Jon A. and Roberts, Sean T.},
  year = {2016},
  journal = {The Journal of Physical Chemistry Letters},
  shortjournal = {J. Phys. Chem. Lett.},
  volume = {7},
  number = {23},
  pages = {4922--4928},
  publisher = {American Chemical Society},
  doi = {10.1021/acs.jpclett.6b02320},
  url = {https://doi.org/10.1021/acs.jpclett.6b02320},
  urlaccess = {2025-10-24}
}

@article{shengActiveControlElectromagnetic2015,
  title = {Active Control of Electromagnetic Radiation through an Enhanced Thermo-Optic Effect},
  author = {Sheng, Chong and Liu, Hui and Zhu, Shining and Genov, Dentcho A.},
  year = {2015},
  journal = {Scientific Reports},
  shortjournal = {Sci Rep},
  volume = {5},
  number = {1},
  pages = {8835},
  publisher = {Nature Publishing Group},
  issn = {2045-2322},
  doi = {10.1038/srep08835},
  url = {https://www.nature.com/articles/srep08835},
  urlaccess = {2025-10-24},
  langid = {english}
}

@article{pensackExcitonDelocalizationDrives2015,
  title = {Exciton {{Delocalization Drives Rapid Singlet Fission}} in {{Nanoparticles}} of {{Acene Derivatives}}},
  author = {Pensack, Ryan D. and Tilley, Andrew J. and Parkin, Sean R. and Lee, Tia S. and Payne, Marcia M. and Gao, Dong and Jahnke, Ashlee A. and Oblinsky, Daniel G. and Li, Peng-Fei and Anthony, John E. and Seferos, Dwight S. and Scholes, Gregory D.},
  year = {2015},
  journal = {Journal of the American Chemical Society},
  shortjournal = {J. Am. Chem. Soc.},
  volume = {137},
  number = {21},
  pages = {6790--6803},
  publisher = {American Chemical Society},
  issn = {0002-7863},
  doi = {10.1021/ja512668r},
  url = {https://doi.org/10.1021/ja512668r},
  urlaccess = {2025-10-27}
}

@article{paceDynamicsSingletFission2018,
  title = {Dynamics of Singlet Fission and Electron Injection in Self-Assembled Acene Monolayers on Titanium Dioxide},
  author = {Pace, Natalie A. and Arias, Dylan H. and Granger, Devin B. and Christensen, Steven and Anthony, John E. and Johnson, Justin C.},
  year = {2018},
  journal = {Chemical Science},
  shortjournal = {Chem. Sci.},
  volume = {9},
  number = {11},
  pages = {3004--3013},
  publisher = {The Royal Society of Chemistry},
  issn = {2041-6539},
  doi = {10.1039/C7SC04688J},
  url = {https://pubs.rsc.org/en/content/articlelanding/2018/sc/c7sc04688j},
  urlaccess = {2025-10-27},
  langid = {english}
}

@article{qiuElectronBeamEvaporated2015,
  title = {An Electron Beam Evaporated {{TiO2}} Layer for High Efficiency Planar Perovskite Solar Cells on Flexible Polyethylene Terephthalate Substrates},
  author = {Qiu, Weiming and Paetzold, Ulrich W. and Gehlhaar, Robert and Smirnov, Vladimir and Boyen, Hans-Gerd and Tait, Jeffrey G. and Conings, Bert and Zhang, Weimin and Nielsen, Christian B. and McCulloch, Iain and Froyen, Ludo and Heremans, Paul and Cheyns, David},
  year = {2015},
  journal = {Journal of Materials Chemistry A},
  shortjournal = {J. Mater. Chem. A},
  volume = {3},
  number = {45},
  pages = {22824--22829},
  publisher = {The Royal Society of Chemistry},
  issn = {2050-7496},
  doi = {10.1039/C5TA07515G},
  url = {https://pubs.rsc.org/en/content/articlelanding/2015/ta/c5ta07515g},
  urlaccess = {2025-10-28},
  langid = {english}
}

@article{bualiIntersystemCrossingOutcompetes2025,
  title = {Intersystem {{Crossing Outcompetes Triplet-Pair Separation}} from 1({{TT}}) below 270 {{K}} in {{Anthradithiophene Films}}},
  author = {Bu Ali, Eman M. and Bertran, Arnau and Moise, Gabriel and Wang, Shuangqing and Kilbride, Rachel C. and Anthony, John E. and Tait, Claudia E. and Clark, Jenny},
  year = {2025},
  journal = {Journal of the American Chemical Society},
  shortjournal = {J. Am. Chem. Soc.},
  volume = {147},
  number = {32},
  pages = {28638--28650},
  publisher = {American Chemical Society},
  issn = {0002-7863},
  doi = {10.1021/jacs.5c00001},
  url = {https://doi.org/10.1021/jacs.5c00001},
  urlaccess = {2025-11-05}
}

@article{wolfQuantumBeatsMultiexciton2018,
  title = {Quantum Beats of a Multiexciton State in Rubrene Single Crystals},
  author = {Wolf, Eric A. and Finton, Drew M. and Zoutenbier, Vincent and Biaggio, Ivan},
  year = {2018},
  journal = {Applied Physics Letters},
  shortjournal = {Appl. Phys. Lett.},
  volume = {112},
  number = {8},
  pages = {083301},
  issn = {0003-6951},
  doi = {10.1063/1.5020652},
  url = {https://doi.org/10.1063/1.5020652},
  urlaccess = {2025-11-07}
}

@article{ryasnyanskiyTripletExcitonDynamics2011a,
  title = {Triplet Exciton Dynamics in Rubrene Single Crystals},
  author = {Ryasnyanskiy, Aleksandr and Biaggio, Ivan},
  year = {2011},
  journal = {Physical Review B},
  shortjournal = {Phys. Rev. B},
  volume = {84},
  number = {19},
  pages = {193203},
  publisher = {American Physical Society},
  doi = {10.1103/PhysRevB.84.193203},
  url = {https://link.aps.org/doi/10.1103/PhysRevB.84.193203},
  urlaccess = {2025-11-07}
}

@article{herreraMolecularPolaritonsControlling2020a,
  title = {Molecular Polaritons for Controlling Chemistry with Quantum Optics},
  author = {Herrera, Felipe and Owrutsky, Jeffrey},
  year = {2020},
  journal = {The Journal of Chemical Physics},
  shortjournal = {J. Chem. Phys.},
  volume = {152},
  number = {10},
  pages = {100902},
  issn = {0021-9606},
  doi = {10.1063/1.5136320},
  url = {https://doi.org/10.1063/1.5136320},
  urlaccess = {2025-11-18}
}

@article{siltanenEnhancingEfficiencyPolariton2025a,
  title = {Enhancing the {{Efficiency}} of {{Polariton OLEDs}} in and {{Beyond}} the {{Single-Excitation Subspace}}},
  author = {Siltanen, Olli and Luoma, Kimmo and Musser, Andrew J. and Daskalakis, Konstantinos S.},
  year = {2025},
  journal = {Advanced Optical Materials},
  volume = {13},
  number = {12},
  pages = {2403046},
  issn = {2195-1071},
  doi = {10.1002/adom.202403046},
  url = {https://onlinelibrary.wiley.com/doi/abs/10.1002/adom.202403046},
  urlaccess = {2025-11-25},
  langid = {english}
}

@article{georgiouGenerationAntiStokesFluorescence2018a,
  title = {Generation of {{Anti-Stokes Fluorescence}} in a {{Strongly Coupled Organic Semiconductor Microcavity}}},
  author = {Georgiou, Kyriacos and Jayaprakash, Rahul and Askitopoulos, Alexis and Coles, David M. and Lagoudakis, Pavlos G. and Lidzey, David G.},
  year = {2018},
  journal = {ACS Photonics},
  shortjournal = {ACS Photonics},
  volume = {5},
  number = {11},
  pages = {4343--4351},
  publisher = {American Chemical Society},
  doi = {10.1021/acsphotonics.8b00552},
  url = {https://doi.org/10.1021/acsphotonics.8b00552},
  urlaccess = {2026-01-07}
}

@article{colesImagingPolaritonRelaxation2013,
  title = {Imaging the Polariton Relaxation Bottleneck in Strongly Coupled Organic Semiconductor Microcavities},
  author = {Coles, David M.},
  year = {2013},
  journal = {Physical Review B},
  shortjournal = {Phys. Rev. B},
  volume = {88},
  number = {12},
  doi = {10.1103/PhysRevB.88.121303}
}

@article{tayebjeeQuintetMultiexcitonDynamics2017,
  title = {Quintet Multiexciton Dynamics in Singlet Fission},
  author = {Tayebjee, Murad J. Y. and Sanders, Samuel N. and Kumarasamy, Elango and Campos, Luis M. and Sfeir, Matthew Y. and McCamey, Dane R.},
  year = {2017},
  journal = {Nature Physics},
  shortjournal = {Nature Phys},
  volume = {13},
  number = {2},
  pages = {182--188},
  publisher = {Nature Publishing Group},
  issn = {1745-2481},
  doi = {10.1038/nphys3909},
  url = {https://www.nature.com/articles/nphys3909},
  urlaccess = {2026-01-08},
  langid = {english}
}

@article{weissStronglyExchangecoupledTriplet2017,
  title = {Strongly Exchange-Coupled Triplet Pairs in an Organic Semiconductor},
  author = {Weiss, Leah R. and Bayliss, Sam L. and Kraffert, Felix and Thorley, Karl J. and Anthony, John E. and Bittl, Robert and Friend, Richard H. and Rao, Akshay and Greenham, Neil C. and Behrends, Jan},
  year = {2017},
  journal = {Nature Physics},
  shortjournal = {Nature Phys},
  volume = {13},
  number = {2},
  pages = {176--181},
  publisher = {Nature Publishing Group},
  issn = {1745-2481},
  doi = {10.1038/nphys3908},
  url = {https://www.nature.com/articles/nphys3908},
  urlaccess = {2026-01-08},
  langid = {english}
}

@article{taitComputationalToolsSimulation2023,
  title = {Computational Tools for the Simulation and Analysis of Spin-Polarized {{EPR}} Spectra},
  author = {Tait, Claudia E. and Krzyaniak, Matthew D. and Stoll, Stefan},
  year = {2023},
  journal = {Journal of Magnetic Resonance},
  shortjournal = {Journal of Magnetic Resonance},
  volume = {349},
  pages = {107410},
  issn = {1090-7807},
  doi = {10.1016/j.jmr.2023.107410},
  url = {https://www.sciencedirect.com/science/article/pii/S1090780723000459},
  urlaccess = {2026-01-08}
}

@article{polakManipulatingMoleculesStrong2020,
  title = {Manipulating Molecules with Strong Coupling: Harvesting Triplet Excitons in Organic Exciton Microcavities},
  shorttitle = {Manipulating Molecules with Strong Coupling},
  author = {Polak, Daniel and Jayaprakash, Rahul and Lyons, Thomas P. and Martínez-Martínez, Luis Á and Leventis, Anastasia and Fallon, Kealan J. and Coulthard, Harriet and Bossanyi, David G. and Georgiou, Kyriacos and Anthony J. Petty, I. I. and Anthony, John and Bronstein, Hugo and Yuen-Zhou, Joel and Tartakovskii, Alexander I. and Clark, Jenny and Musser, Andrew J.},
  year = {2020},
  journal = {Chemical Science},
  shortjournal = {Chem. Sci.},
  volume = {11},
  number = {2},
  pages = {343--354},
  publisher = {The Royal Society of Chemistry},
  issn = {2041-6539},
  doi = {10.1039/C9SC04950A},
  url = {https://pubs.rsc.org/en/content/articlelanding/2020/sc/c9sc04950a},
  urlaccess = {2026-01-29},
  langid = {english}
}

@article{ashiotisFastAzimuthalIntegration2015,
  title = {The Fast Azimuthal Integration {{Python}} Library: {{pyFAI}}},
  shorttitle = {The Fast Azimuthal Integration {{Python}} Library},
  author = {Ashiotis, Giannis and Deschildre, Aurore and Nawaz, Zubair and Wright, Jonathan P. and Karkoulis, Dimitrios and Picca, Frédéric Emmanuel and Kieffer, Jérôme},
  year = {2015},
  journal = {Journal of Applied Crystallography},
  shortjournal = {J Appl Crystallogr},
  volume = {48},
  eprint = {25844080},
  eprinttype = {pubmed},
  pages = {510--519},
  issn = {0021-8898},
  doi = {10.1107/S1600576715004306},
  issue = {Pt 2},
  langid = {english},
  pmcid = {PMC4379438}
}

\end{document}